\documentclass[reprint, aps, prd,superscriptaddress, nofootinbib]{revtex4-1} %
\usepackage{graphicx}%
\usepackage{dcolumn}%
\usepackage{bm}%
\usepackage[linktocpage,colorlinks,pdfusetitle,allcolors=blue]{hyperref}
\usepackage{amsmath, amssymb}
\usepackage{siunitx}
\usepackage{multirow}
\usepackage{booktabs}

\renewcommand{\d}[1]{\ensuremath{\operatorname{d}\!{#1}}}
\renewcommand{\vec}[1]{\boldsymbol{#1}}
\newcommand{\conj}[1]{{#1}^*}
\DeclareSIUnit[]\admmass{\text{\ensuremath{M}}}
\DeclareSIUnit[]\sunmass{\text{\ensuremath{M_{\odot}}}}
\DeclareSIUnit[]\erg{\;\text{erg}\;}

\usepackage{mathtools}
\DeclarePairedDelimiter\abs{\lvert}{\rvert}

\begin{document}

\title{Numerical-relativity simulations of the quasi-circular inspiral and
  merger of non-spinning, charged black holes: methods and comparison with
  approximate approaches} %

\author{Gabriele Bozzola}
\email{gabrielebozzola@arizona.edu}
\affiliation{Department of Astronomy, University of Arizona, Tucson, AZ, USA}
\author{Vasileios Paschalidis}
\email{vpaschal@arizona.edu}
\affiliation{Department of Astronomy, University of Arizona, Tucson, AZ, USA}
\affiliation{Department of Physics, University of Arizona, Tucson, AZ, USA}

\date{\today}

\begin{abstract}
  We present fully general relativistic simulations of the quasi-circular
  inspiral and merger of charged, non-spinning, binary black holes with
  charge-to-mass ratio $\lambda \le \num{0.3}$. We discuss the key features that
  enabled long term and stable evolutions of these binaries. We also present a
  formalism for computing the angular momentum carried away by electromagnetic
  waves, and the electromagnetic contribution to black-hole horizon properties.
  We implement our formalism and present the results for the first time in
  numerical-relativity simulations. In addition, we compare our full non-linear
  solutions with existing approximate models for the inspiral and ringdown
  phases. We show that Newtonian models based on the quadrupole approximation
  have errors of \SI{20}{\percent}--\SI{100}{\percent} in key gauge-invariant
  quantities. On the other hand, for the systems considered, we find that
  estimates of the remnant black hole spin based on the motion of test particles
  in Kerr-Newman spacetimes agree with our non-linear calculations to within a
  few percent.  Finally, we discuss the prospects for detecting black hole
  charge by future gravitational-wave detectors using either the
  inspiral-merger-ringdown signal or the ringdown signal alone.
\end{abstract}

\maketitle

\section{Introduction}
\label{sec:introduction}

In previous work~\cite{Bozzola2019, Bozzola2020}, we initiated a systematic
program to study the interactions between charged black holes in full non-linear
Einstein-Maxwell theory. These explorations are relevant for gravitational-wave
astronomy, exotic astrophysics, and fundamental physics (such as modified
gravity and beyond-standard-model physics). We first review these applications
in Sec.~\ref{sec:inter-dynam-stud}. We summarize the main goals and results of
this work in Sec.~\ref{sec:goals-this-work}, and outline the structure of the
manuscript and conventions adopted in Sec.~\ref{sec:struct-paper-conv}. Readers
that are mainly interested in our new results can skip
Sec.~\ref{sec:inter-dynam-stud}.

\subsection{Motivation for non-linear simulations in Einstein-Maxwell theory}
\label{sec:inter-dynam-stud}

Over the past few years there has been a growing interest in modified theories
of gravity to perform strong-field tests of general relativity. The data
collected by the Event Horizon Telescope~\cite{EHT1} allowed for new tests of
gravity around supermassive black holes, and the observation of gravitational
waves by the LIGO-Virgo collaboration~\cite{LIGOVirgo2018} enabled the first
constraints on deviations from Einstein's theory in the highly dynamical
strong-field regime (see, e.g.~\cite{Abbott2016o, Yagi2016, Abbott2019b,
  Abbott2019c}). In turn, this latter achievement was made possible by
advancements in the field of numerical relativity that has been able to produce
the accurate gravitational-wave models that are needed to detect these signals
and perform the associated parameter estimations. Therefore, it should come to
no surprise that the scarcity of simulations of compact binary mergers in
modified theories of gravity~\cite{Yunes:2016jcc} severely limits out our
ability to use gravitational-wave observations to place stronger constraints.
The reason for this shortage is that many models of modified gravity are
``sick'' (e.g., lack of well-posedness, ghosts, etc., see~\cite{Cayuso:2017iqc}
for a discussion), thereby making these computations particularly challenging or
impossible. Hence, with the exception of a few
cases~\cite{Hirschmann:2017psw,Ripley:2020vpk,East:2020hgw}, progress in this
direction is usually made by means of order-reduced approaches, and not
solutions of the full theory~(see, e.g.~\cite{Okounkova2017, Okounkova2020,
  Witek2019, Okounkova2020b, Witek2020}). In contrast to modified gravity,
Einstein-Maxwell theory admits a well-posed initial-value problem, while sharing
other non-trivial properties with modified gravity (e.g, emission of dipole
radiation\footnote{We note that this is not gravitational dipole radiation.}).
Moreover, some modified gravity theories reduce effectively to Einstein-Maxwell
in specific limits (e.g.~\cite{Moffat2006}). Therefore, the inspiral and merger
of charged black holes constitutes, in a sense, the middle ground between
traditional general relativity and modified theories of gravity. Despite these
facts, the non-linear dynamics of charged binary black holes is uncharted
territory, with~\cite{Zilhao2012,Zilhao2013,Zilhao2014,Zilhao2015,Liebling2016}
being the main works on the subject.

Studying charged black holes in Einstein-Maxwell theory not only provides a way
to capture some features of specific modified theories in a controlled
environment, but also simulations of such systems have direct astrophysical and
fundamental physics applications. First, while black holes are expected to be
electrically neutral~\cite{Wald1974,Gibbons1975,Eardley1975,Gong2019,Pan2019},
there is no definitive observational support for this expectation. Therefore,
this assumption must be tested. Gravitational-wave observations offer a
model-independent way to test this assumption. Second, ``charge'' is an umbrella
term that applies to different models in exotic astrophysics and
beyond-standard-model physics, including dark matter with hidden charge and
interacting with dark electromagnetism
(e.g.~\cite{Feng2009,Ackerman2009,Foot2015,Foot2015b,
  Foot2016,Agrawal:2016quu,Christiansen2020} or dark matter with fractional
charge~\cite{Davidson1991,Perl1997,Davidson2000,Dubovsky2004,Chatrchyan2013,Dolgov2013,
  Vogel2014,Cardoso2016b,Gautham2019,Plestid2020}), as well as modified theories
of gravity with additional vector fields sourced by a ``gravitational''
charge~\cite{Moffat2006}. Furthermore, through a duality transformation,
``charge'' can also be interpreted as \emph{magnetic charge}. Recently, black
holes with magnetic charge in astrophysics have received some
attention~(e.g.~\cite{Liu2020, Liu2020b, Liu2020c, Bai2020, Ghosh2020}), with
focus on primordial black holes in~\cite{Liu2020,Liu2020b}. Gravitational waves
can be used to directly test whether black holes have any kind of charge. This
was the goal of our previous paper on the subject~\cite{Bozzola2020}, where we
found that charge-to-mass ratios of up to $\lambda = 0.3$\footnote{We are using
  geometrized units. See Sec.~\ref{sec:struct-paper-conv}} are compatible with
GW150914~\cite{Abbott2016}, assuming that the role of black hole spins can be
neglected. Upper bounds on black hole charge can be translated to constraints on
the properties of the dark matter particles in the aforementioned models or on
the parameters of modified gravity theories~\cite{Bozzola2020}.

Numerical-relativity simulations of charged black holes can be used to produce
gravitational-wave templates that include charge (for example, by
\emph{hybridizing} analytical waveforms with numerical ones, as done
in~\cite{Hannam2008, MacDonald2011, Ajith2012, Aasi2014, Mehta2017, Varma2019}).
Since the detection of gravitational waves by the LIGO-Virgo interferometers
relies heavily on matched-filtering
techniques~\cite{Sathyaprakash1991,Dhurandhar1994, Balasubramanian1996},
extended gravitational-wave template banks that encode additional physics are
necessary for the parameter estimation~\cite{Flanagan1998a, Flanagan1998b,
  Aylott2009, Ajith2012, Hinder:2013oqa}. A phenomenological model based on
numerical relativity with charge can also be used in Bayesian analyses to
directly constrain this parameter in LIGO-Virgo signals.

Finally, charge provides a way for a black hole to reach extremality (along with
the spin). Thus, non-linear studies of charged binary black holes offer new
pathways to investigate cosmic censorship in conditions where it has never been
probed before. For instance, it would be of interest to tackle the question:
``can black holes be overcharged?'', going beyond previous perturbative
approaches~\cite{Hubeny1999, Wald2018}.

\subsection{Goals of this work}
\label{sec:goals-this-work}

In this paper, we continue our explorations of the non-linear interaction of
charged black holes by approaching the problem of inspirals and mergers on two
different thrusts. On one side, we present necessary ingredients for performing
long-term and stable numerical relativity simulations of the quasi-circular
inspiral of charged binary black holes. In particular, we discuss how
Kreiss-Oliger dissipation~\cite{Kreiss:1973aa} helps (or impedes) these
evolutions. We also describe the formalism that we adopt for our evolutions
detailing some features that have not been included in previous works (e.g., the
computation of the angular momentum carried away by electromagnetic waves with
the Newman-Penrose formalism, and the contribution of electromagnetic fields to
the quasi-local spin of a black hole). Using the gravitational waveforms
generated by our simulations, we explore black hole charge detectability by
future ground- and space-based gravitational wave detectors. The second thrust
of this work consists of analyzing existing approximate models for the inspiral
of charged black hole binaries and their remnant black holes, and comparing them
with our non-linear solutions.

In~\cite{Bozzola2020}, we presented the first simulations of the quasi-circular
inspiral and merger of charged black holes in full general relativity with valid
initial data. Here, we present more details about these computations. We focus
on systems with mass ratio $q=29\slash36$, as inferred for
GW150914~\cite{Abbott2016,Abbott2016d}, and restrict the charge-to-mass ratio
$\lambda$ of the individual black holes to values $|\lambda|\le0.3$. The mass ratio is close
to unity, so we expect that the conclusions presented in this work will hold for
equal-mass binaries. We consider three systems: (1) binary black holes with same
charge-to-mass ratio in magnitude and sign, (2) binary black holes with
charge-to-mass ratio equal in magnitude but oppositely charged, and (3) binary
black holes in which only the primary is charged. In this first exploration, we
do not study cases where the black holes have different (non-zero)
charge-to-mass ratio $\lambda$.

For the comparison of our solutions with existing approximations for the
inspiral phase, we consider a model that is based on Newtonian physics coupled
with the quadrupole formula to incorporate radiation reaction. We will refer to
this model with the letters ``QA'' (\emph{quadrupole approximation}). Given its
simplicity, this model has been routinely used to study the merger of charged
black holes
(e.g.~\cite{Cardoso2016b,Wang2020,Christiansen2020,Liu2020,Liu2020b,Liu2020c,Cardoso2020Erratum}).
Prior work on head-on collisions of charged black hole reported good agreement
in some quantities between these approximate calculations and full non-linear
simulations~\cite{Zilhao2012,Zilhao2013}. Therefore, a goal of this work is to
determine the errors of the Newtonian approximation when applied to the
quasi-circular inspiral of charged black holes. A key result of our study is
that Newtonian models can be successfully applied to obtain order-of-magnitude
estimates of observables or to build intuition, but they cannot be used for
precision studies of these mergers.

A second focus of this paper is on the properties of the post-merger black hole
and its quasi-normal modes. This is especially relevant for LISA, which will
detect the ringdown signal arising from the merger of supermassive black holes
with high signal-to-noise ratio~\cite{Amaro2017}. The quasi-normal modes can be
used to test general relativity~\cite{Berti2006}, as their characteristic
frequencies depend on the spin, mass, and charge of the remnant black hole in a
known way. Here, we consider the method described in~\cite{Jaiakson2017} to
estimate the remnant black hole spin using conservation arguments, and we
compare it with our non-linear solutions. As we discuss later, we find that the
quasi-normal-mode properties do not change much across the simulations
considered here. This is due to the value of the final spin and to the
relatively weak dependence of the quasi-normal-modes on the charge for the
values of $\lambda$ we consider.

\subsection{Structure of the paper and conventions}
\label{sec:struct-paper-conv}

The structure of the remainder of the paper is the following. In
Sec.~\ref{sec:numerical-relativity}, we describe the formalism that we adopt to
perform simulations of the quasi-circular inspiral and merger of charged black
holes. In particular, we discuss how to obtain stable quasi-circular inspirals,
and highlight some new features of the approach. Next
(Sec.~\ref{sec:newtonian-model}), we outline the simplest (Newtonian) model for
the quasi-circular inspiral of non-spinning, charged binary black holes.
Sec.~\ref{sec:final-black-hole} describes what we can say about the remnant
black hole using results of relativistic calculations and perturbation theory.
In Sec.~\ref{sec:results}, we follow the black holes through their coalescence:
first (Sec.~\ref{sec:inspiral}) we study the inspiral and compare the non-linear
solution with the Newtonian model, then (Sec.~\ref{sec:up-merger}) we discuss
results from the full simulations, and finally (Sec.~\ref{sec:prop-final-black})
we report on the properties of the remnant black hole. Conclusions and future
directions are collected in Sec.~\ref{sec:conclusion}.

We adopt geometrized units with $G = c = 1$, with $G$ being Newton's constant
and $c$ the speed of light in vacuum. We also adopt Gaussian units for the
electromagnetic sector. Similarly, we denote the different simulations adding a
superscript and a subscript when we report physical quantities. For example,
$e^{+}_{+}$, $e^{+}_{-}$, and $e^{+}_{{{\kern 0.12em}0}}$ indicate the
eccentricity measured in the evolutions where black holes have charges with the
same signs, with opposite signs, and only one charged black hole,
respectively.\footnote{For figures reporting different charge-to-mass ratios, we
  use a consistent style: red dashed lines with circles are for systems with
  both black holes charged with the same sign, blue dotted lines with squares
  are for oppositely charged ones, and green dash dotted lines with triangles
  are for evolutions in which only one black hole is charged.} In geometrized
units, quantities have units of length. Here we report all the results in units
of the Arnowitt-Deser-Misner (ADM) mass of the system
\si{\admmass}~\cite{Arnowitt2008}. We use the letters $a,b,c,d$ for spacetime
indices, and $i,j,k$ for spatial ones. For everything else, we follow the same
conventions as in~\cite{MTW1973}.

\section{Methods and formalism}
\label{sec:numerical-relativity}

In this section, we describe the methods we adopt for solving the full
non-linear Einstein-Maxwell equations (Sec.~\ref{sec:equat-numer-setup}). We
discuss our approach to building quasi-circular initial data
(Sec.~\ref{sec:contr-eccentr}) and how to achieve long-term, stable evolutions
(Sec.~\ref{sec:achieving-long-term}). We also describe the formalism and
implementation of two new features that have not appeared in previous
simulations: the contribution of the electromagnetic fields to black-hole
horizon properties (Sec.~\ref{sec:contribution-spin-em}), and the angular
momentum carried away by electromagnetic waves
(Sec.~\ref{sec:grav-electr-waves}).

\subsection{Equations and numerical setup}
\label{sec:equat-numer-setup}

In this paper, we study systems described by the source-free Einstein-Maxwell
equations~\cite{Wald1984} (electrovacuum\footnote{In all our discussion, we
  assume that the black holes are in vacuum (see~\cite{Cardoso2020} for a
  discussion on the role of the environment). We also ignore Schwinger
  pair-production and any other quantum effects.})
\begin{subequations}
  \label{eq:einstein-equations}
  \begin{align}
    \mathcal{R}_{ab} - \frac{1}{2}g_{ab} \mathcal{R}  &=
     8 \pi T^{\mathrm{EM}}_{ab}   \,, \\
    \nabla_a F^{ab} &= 0\,, \\
    \nabla_a {}^{\star}F^{ab} &= 0 \,,
  \end{align}
\end{subequations}
where $\mathcal{R}_{ab}$ is the Ricci tensor associated with the metric
$g_{ab}$, $\mathcal{R}=\mathcal{R}^{a}_{\; a}$, $F_{ab} = 2\, A_{[a,b]}$ is the
Maxwell field-strength tensor, with $A_a$ the electromagnetic four-vector
potential, and ${}^{\star}F_{ab}$ is its Hodge dual, defined by
\begin{equation}
\label{eq:fstar}
{}^{\star}F^{ab} = \frac{1}{2} \epsilon^{abcd} \,F_{cd}\,,
\end{equation}
with $\epsilon^{abcd}$ being the Levi-Civita tensor. The electromagnetic
stress-energy tensor is given by
\begin{equation}
  \label{eq:electromagnetic-stress-energy-tensor}
  4 \pi T_{ab}^{\mathrm{EM}} =  F_{ac} F_{bd} g^{cd} -
  \frac{1}{4}g_{ab} F_{cd} F^{cd}\,.
\end{equation}

We solve the coupled Einstein-Maxwell equations in a $3+1$ decomposition of the
spacetime (for more details, see Sec. II A in~\cite{Bozzola2019}, or textbooks
on the subject, e.g.~\cite{Alcubierre:2008it,Baumgarte:2010nu,Shibata2016b}) and
use the \texttt{Einstein
  Toolkit}~\cite{Loffler:2011ay,EinsteinToolkit:ascl,EinsteinToolkit:web} for
the numerical integration.

The initial data are generated by \texttt{TwoChargedPunctures}, which solves the
Hamiltonian constraint equation using a Bowen-York approach~\cite{Bowen:1980yu,
  Bowen1985}. We developed and tested \texttt{TwoChargedPunctures}
in~\cite{Bozzola2019}, starting from the widely used \texttt{TwoPunctures}
code~\cite{Ansorg2004}. \texttt{TwoChargedPunctures} takes as input the
locations, charges, bare masses, angular, and linear momenta of each of the two
black holes, and it can build arbitrary configurations. We start our simulations
at a coordinate distance of \SI{12.1}{\admmass}, with the two black holes having
fixed charge-to-mass ratio $\lambda$ and mass-ratio $29\slash36$. In this first
study, we only explore systems with the two black holes having the same
$\lambda$ (up to the sign), or with only one charged black hole. In
Sec.~\ref{sec:contr-eccentr}, we discuss how to choose the initial data
parameters to achieve quasi-circular inspirals.

The evolution of the spacetime is performed with the \texttt{Lean}
code~\cite{Sperhake2007}, which implements the
Baumgarte-Shapiro-Shibata-Nakamura (BSSN) formulation of Einstein's
equations~\cite{Shibata1995,Baumgarte1998}. \texttt{Lean} evolves the conformal
factor $\chi_{\text{conf}} = \gamma^{-1\slash6}$, with $\gamma$ determinant of the
3-metric. We adopt as spacetime gauges the $1 + \log$ and the
$\Gamma-\text{driver}$ conditions~\cite{Alcubierre:2000xu,
  Alcubierre:2002kk}. The electromagnetic fields are evolved with the massless
version of the \texttt{ProcaEvolve}~\cite{Zilhao2015} code. The code evolves the
electric field $E^{i}$, the scalar and vector potentials $\phi$ and $A_{i}$,
along with an auxiliary variable $Z$ that is used to control the Maxwell
constraints. The precise equations solved by \texttt{ProcaEvolve} are presented
in the Appendix of~\cite{Zilhao2015}. These codes are publicly available as part
of the Canuda suite~\cite{canuda,canudacode} and have been extensively tested
and used throughout the years. We use sixth-order accurate finite differences
for the spatial derivative, and we integrate in time with a fourth-order
Runge-Kutta method.

Apparent horizons are located using
\texttt{AHFinderDirect}~\cite{Thornburg:1995cp,Thornburg:2003sf}, and their
physical properties are measured with \texttt{QuasiLocalMeasuresEM}, a version
of \texttt{QuasiLocalMeasures}~\cite{Dreyer:2002mx} updated to implement the
isolated horizon formalism in full Einstein-Maxwell theory (see Sec.~II~C
in~\cite{Bozzola2019}). This extension is necessary when considering black holes
in the presence of electromagnetic fields, as we discuss in
Sec.~\ref{sec:contribution-spin-em}.

We extract waves via the Newman-Penrose formalism as implemented in
\texttt{NPScalars\_Proca} (see, Sec.~\ref{sec:grav-electr-waves}) and recover
the gravitational-wave strain with a time integration using the fixed frequency
integration method~\cite{Reisswig2011}. We consider a finite extraction radius
of \SI{110.69}{\admmass}. Results are approximately invariant if we consider
different extraction radii or if we extrapolate the waves to infinity with the
method described in~\cite{Hinder:2013oqa}.

We work with Cartesian grids with Berger-Oliger adaptive mesh refinement as
provided by \texttt{Carpet}~\cite{Schnetter:2003rb}. We use two sets of nine
nested refinement levels that are centered on and track the centroid of the
black hole apparent horizons. The resolution of our simulations is $M\slash65$,
where $M$ is the total ADM mass, with additional resolutions to perform
convergence study (which we reported in~\cite{Bozzola2020}). The outer boundary
is at \SI{1033}{\admmass} and we performed selected simulations to verify that
the location does not affect the evolution. Some more details on evolution and
grid parameters used in our simulations are reported in
Appendix~\ref{sec:param-test-evol}.

\subsection{Controlling the eccentricity}
\label{sec:contr-eccentr}

In this work, we focus on quasi-circular inspirals. When considering charged
black holes, the effect of the electromagnetic fields must be taken into account
to achieve a low-eccentricity coalescence. Here, we describe a simple method to
incorporate the effect of charge. This approach successfully yields
quasi-circular inspirals for the values of $\lambda$ explored in this study.

First, it is useful to summarize how quasi-circular inspirals are obtained in
the case without charge. The simplest way is to start from Newtonian physics.
Consider two point particles with mass $m_1$, $m_2$, and assume that they are in
a circular Keplerian orbit. The orbital angular velocity $\Omega$ of each
particle is (restoring the gravitational constant $G$)
\begin{equation}
  \label{eq:ang-vel-orbit}
  \Omega = \sqrt{\frac{G(m_{1} + m_{2})}{d^{3}}}\,,
\end{equation}
where $m_i$ is the mass of the $i$-th component, and $d$ the orbital
separation. If we denote the mass-ratio $q=m_1/m_2$, the linear velocity of two
particles becomes
\begin{equation}
  \label{eq:ang-vel-orbit-lin}
  v_{1} = \frac{\Omega d}{1+q} \quad \text{and} \quad v_{2} = q \frac{\Omega d}{1+q}\,.
\end{equation}
In numerical integrations, black holes are assumed to behave like these point
masses, so $p_i=m_i v_{i}$ is the initial linear momentum assigned to the $i$-th
black hole. Evolutions initialized with such Newtonian values have significant
residual eccentricity~\cite{Pfeiffer:2007yz}, hence high-order post-Newtonian
(PN) expansions are used to compute more accurately the linear momenta necessary
for quasi-circularity. When going beyond the Newtonian approximation, radial
contribution to the velocities appear.

Now, let us endow the point particles with charges $q_1 = \lambda_1 m_1$, and
$q_2 = \lambda_2 m_2$ (with $\lambda$ being the charge-to-mass ratio). In
Newtonian physics, both electromagnetism and gravity are central forces, so,
from the point of view of the dynamics, this system is indistinguishable from
one with uncharged bodies but gravitational constant $\widetilde{G} = (1 -
\lambda_1 \lambda_2)G$. For this reason, one can incorporate the effect of
charge by rescaling $G$ to $\widetilde{G}$. We can use this fact to achieve
low-eccentricity inspirals for charged black holes. First, we compute the linear
momenta needed for a quasi-circular coalescence of the black holes without
charges using the highest order post-Newtonian expansion available (in our
simulations we used 2.5PN). Then, we rescale these momenta by $\sqrt{1 -
  \lambda_{1} \lambda_{2}}$ to introduce the effect of electromagnetism. This
simple method is effective at keeping eccentricity under control for the values
of charge-to-mass ratio explored here, as we show next.

\begin{figure}[htbp]
  \centering
  \includegraphics{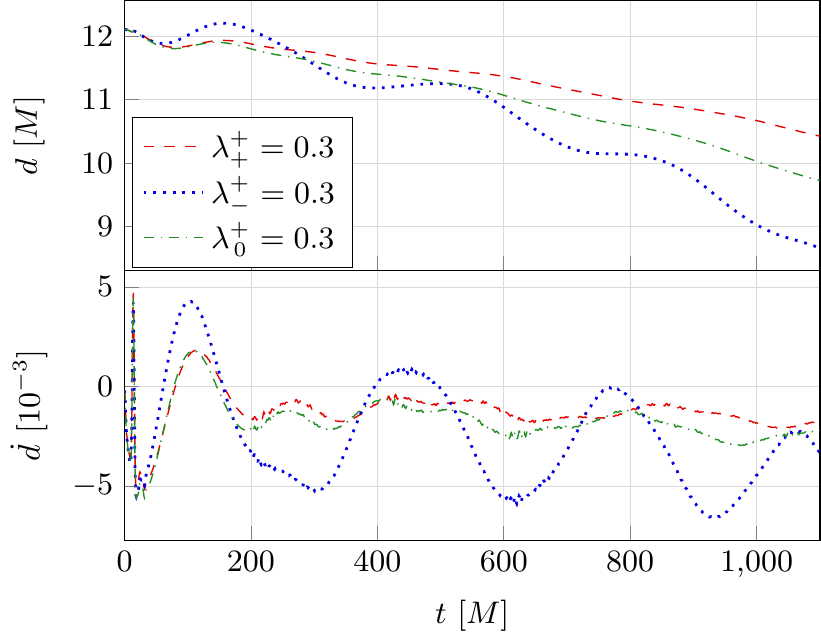}
  \caption{First few orbits for the simulations with charge-to-mass ratio of
    $\lambda = \num{0.3}$. Top panel: (coordinate) separation $d$ of the centroid of
    the two black holes as a function of the (coordinate) time. Bottom panel:
    time derivative of $d$, this quantity can be used to estimate the residual
    eccentricity with Eq.~\eqref{eq:ecc-fit}. The initial spike is due to the
    initial data relaxation, and is not included in our analyses. Note that
    these quantities are not gauge-invariant.}
  \label{fig:eccentricity}
\end{figure}

Following~\cite{Pfeiffer:2007yz}, we estimate the residual eccentricity by
fitting the time derivative of the coordinate separation of the two black holes
$\dot{d}$ with a function of the form
\begin{equation}
  \label{eq:ecc-fit}
  \dot{d}(t) = A_{0} + A_{1} t + B \sin{(\omega t + \varphi)}\,.
\end{equation}
In Fig.~\ref{fig:eccentricity}, we show the coordinate separation $d$ and its
derivative $\dot{d}$ for the most extreme simulations that we are considering
here. The scale of the amplitude of the oscillations in the bottom panel already
provides an idea of the (small) amount of residual eccentricity. The very
beginning of the time series is noisy due to the relaxation of the initial data,
so we exclude that part of the simulation. We perform the fit with the
Levenberg-Marquardt algorithm~\cite{Levenberg1944,Marquardt1963} for non-linear
least-square fitting as implemented in MINPACK\@. As in~\cite{Pfeiffer:2007yz},
we find that fits are not perfect, so the eccentricities reported should be
considered as estimates.  In general, we find that the first orbit is the one
with the most eccentricity.  Fitting the first three orbits we find $e^{+}_{+}
\approx \num{0.01}$, $e^{+}_{-} \approx \num{0.02}$, and $e^{+}_{{{\kern
      0.12em}0}} \approx \num{0.01}$. The eccentricity is significantly reduced
if we consider the next three orbits after the first: $e^{+}_{+} \approx
\num{2e-3}$, $e^{+}_{-} \approx \num{4e-3}$, and $e^{+}_{{{\kern 0.12em}0}}
\approx \num{1e-3}$. 

These values of eccentricity found are remarkably small considering the
simplicity of the method employed. Simulations with higher charge-to-mass ratio,
especially in the case of opposite charges, may have a significant residual
eccentricity. There are at least three ways to improve our method and further
remove the eccentricity.

First, one can adapt iterative eccentricity-reduction schemes, such as those
described in~\cite{Pfeiffer:2007yz, Purrer2008, Ramos-Buades2019} for uncharged
black holes, to systems with charge. Alternatively, one can use post-Newtonian
methods that include directly electromagnetic fields to estimate the linear
momenta. The two methods are not mutually exclusive, and for simulations with
extreme charge both may be required to produce quasi-circular
inspirals. Finally, one can always start the evolution from a larger initial
separation and let gravitational and electromagnetic waves circularize the
orbit.

\subsection{Achieving long-term stable evolutions}
\label{sec:achieving-long-term}

Performing long-term, stable evolutions of black holes in vacuum in 3+1
dimensions used to be the greatest challenge in numerical relativity. Through
substantial developments over the past two decades, solving the Einstein
equations in vacuum is considered a solved problem, and there is considerable
knowledge on the topic. However, simulations of black holes with electromagnetic
fields are still in their infancy. It is not yet clear how much of the
technology developed for vacuum spacetimes carries over to electrovacuum
spacetimes. Being able to perform long-term and stable evolutions of charged
black holes is therefore not granted. Indeed, the simulations presented in this
work are among the longest and most sophisticated to date, and presented some
challenges. We found that adding artificial dissipation to all evolved variables
in a specific way is critical for successful simulations. We found that standard
recipes for artificial dissipation that work in the case of vacuum binary black
hole spacetimes lead to blowups in the case of electrovacuum binary black holes.

Artificial dissipation stabilizes evolutions by removing high-frequency unstable
modes. This approach has proven necessary to achieve long-term simulations in
many cases (see,
e.g.~\cite{Pretorius:2005gq,Baker:2005vv,Campanelli:2005dd,Zlochower2005,
  Calabrese2006, Husa2008, Babiuc2008}). The most common flavor of artificial
dissipation adopted in numerical relativity is known as \emph{Kreiss-Oliger
  dissipation}~\cite{Kreiss:1973aa}. This technique consists of introducing
artificial diffusion by adding a term to the evolution equation of a variable
$U$ as follows (schematically)
\begin{equation}
  \label{eq:diss-1}
  \partial_t U = \cdots + {(-1)}^{(p+3)/2} \frac{\epsilon}{2^{p+1}} \Delta x^{p} \partial^{p+1}_{x}U\,,
\end{equation}
where $\cdots$ indicate the right-hand-side of the evolution equation of $U$, $p$ is
the order of the Kreiss-Oliger dissipation, $\Delta x$ is the grid spacing and the
numerical factor $\epsilon \Delta x^{p} \slash 2^{p+1}$ is the diffusivity, which represents the
strength of the dissipation. Note that although we only add a spatial derivative
in the $x$ direction in Eq.~\eqref{eq:diss-1}, the actual operator has also
corresponding $y$ and $z$ derivatives. For simplicity of the presentation we do
not write these extra terms here. In the infinite resolution limit, this new
term vanishes and the equations are the ones we started with. However, at finite
resolution, it is important to ensure that the modification does not affect the
convergence order of the solution. Thus, $p$ is typically chosen to be greater
than the convergence order of the evolution operator. Moreover, $p$ has to be
odd, so that this modification is an even-order parabolic operator (since $p+1$
is even). In Appendix~\ref{sec:dissipation}, we present some details on
conditions that must be satisfied for numerical stability.

As in previous works, our simulations quickly crash (in the first
\SI{100}{\admmass}) if we do not add artificial diffusion. Since our evolutions
are sixth-order accurate, we add a seventh-order ($p=7$) dissipation to all
evolved variables (i.e., including the spacetime and electromagnetic fields and
the gauge variables). The strength of the dissipation is determined by the
coefficient $\epsilon$ (see, Appendix~\ref{sec:dissipation} for details). We explored
three prescriptions for setting $\epsilon$ on different refinement levels: (1) constant
$\epsilon$, (2) proportional to the local Courant factor $\Delta t/\Delta x$, and (3)
``continuous'' $\epsilon$ (described below). The rationale behind (1) is to have a form
that respects the local Courant stability condition independently of $\Delta x$ as
long as the Courant factor is the same everywhere. It also has increased
effective dissipation in regions with coarser resolution. Numerical stability
requirements often require that different refinement levels be evolved with
different Courant factors, so prescription (2) amends (1) by modifying $\epsilon$ when
the Courant factor is changed to ensure that Courant stability conditions are
met (see Appendix~\ref{sec:dissipation}). While (1) and (2) are commonly used in
numerical relativity, (3) is introduced in this paper and consists of setting
$\epsilon_i = \epsilon_{n} {(\Delta x_i/\Delta x_n)}^{-p}$, where $i=1,2,\ldots,n$ is the index indicating
the refinement level (larger $i$ indicating finer level), with $n$ the maximum
number of refinement levels, and $\Delta x_i$ the grid spacing of refinement level
$i$. The above prescription can also be written as
$\epsilon_i = \epsilon_{0}/{(2^{i-n})}^{p}$, since refinement level grid spacings differ by
factors of 2, i.e., $\Delta x_{i-1}=2\Delta x_i$. Thus, the diffusivity entering
Eq.~\eqref{eq:diss-1} for each level becomes
$\epsilon_{n} {(\Delta x_i/\Delta x_n)}^{-p} \Delta x_i^{p} \slash 2^{p+1} = \epsilon_{n} \Delta x_n^p \slash 2^{p+1}$. In
other words, this new prescription guarantees that the effective diffusivity is
the same everywhere on the grid, ensuring that the same parabolic diffusion
operator is added to the set of equations on each refinement level. Moreover,
the artificial diffusion again goes to 0 at order $p$, and thus does not affect
the expected order of convergence of the finite difference scheme. We call this
\emph{continuous}, because prescriptions (1) and (2) have jumps in the effective
diffusivity across refinement levels that introduce discontinuities in the
equations. Hence, the parabolic operator added to the equations depends on the
refinement level, which results in effectively solving a different system of
partial differential equations on different refinement levels. This difference
vanishes in the limit of infinite resolution. To our knowledge our approach (3)
for setting the Kreiss-Oliger diffusivity on adaptive-mesh-refinement grids has
not been discussed before. We found that this is a crucial ingredient for
long-term and stable numerical evolutions of charged black holes.

\begin{figure}[htbp]
  \centering
  \includegraphics{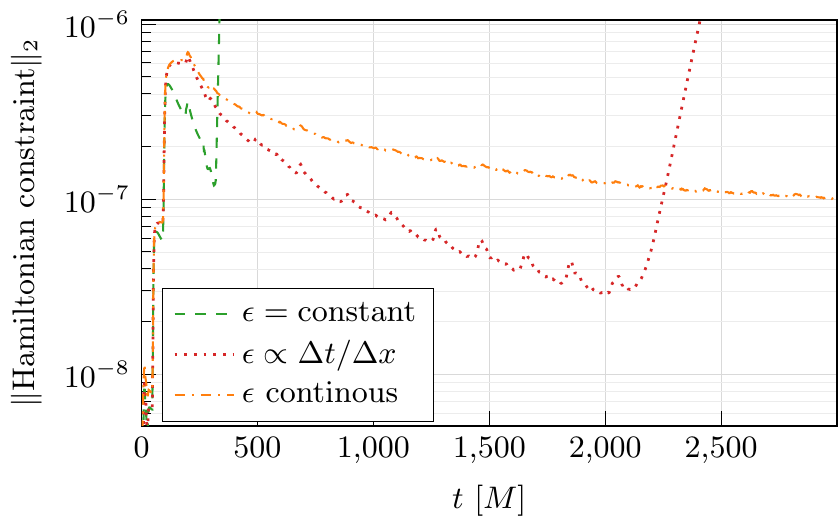}
  \caption{Evolution of the L2 norm of the Hamiltonian constraint (outside the
    horizon) for the three different Kreiss-Oliger dissipation prescriptions.
    The ``continuous'' dissipation prescription introduced in this work is the
    only one that allows long-term and stable simulations with seventh-order
    Kreiss-Oliger dissipation and sixth-order accurate finite differences.}
  \label{fig:hamc}
\end{figure}

To demonstrate the performance of each of the three prescriptions for setting
$\epsilon$, we consider a high-resolution simulation of a single charged,
non-spinning black hole with mass $\si{\admmass}=1$ and charge
$Q = \SI{0.5}{\admmass}$. In Fig.~\ref{fig:hamc}, we show the L2 norm of the
violation of the Hamiltonian constraint (excluding the domain interior to the
apparent horizon) for each dissipation prescription. More details about our
numerical setup are provided in Appendix~\ref{sec:param-test-evol}. The figure
is representative of how our binary charged black hole simulations evolve, and
demonstrates that there is unstable growth in some variables.\footnote{The jumps
  in the constraints at the beginning of the simulations are due to an initial
  pulse in the gauge variables propagating outwards from the center of the
  simulation. Every time this pulse crosses a refinement boundary, it is turned
  into constraint-violating modes through refinement level interpolations. This
  behavior in the constraints is well-known~\cite{Etienne2014}.} We verified
that this instability is numerical and not physical, since its onset depends on
the resolution of the simulation: the higher the resolution, the earlier the
instability takes place. This behavior was not reported in previous studies
without charge, suggesting that the instability first arises in the
electromagnetic sector, and then feeds the gravitational one. Indeed, we observe
the numerical instability starts first in electromagnetic quantities, such as
the Gauss constraint. We also found that the unstable growth of the constraint
first occurs near the outer boundary, suggesting that the approximate
outgoing-wave boundary conditions may be the trigger of the instability.
Finally, we noticed that a fourth-order accurate finite-difference evolution
(with fifth order dissipation) and constant $\epsilon$ does not lead to the same
numerical explosions at the same resolution and within the simulation times we
considered. What is important to note regarding the goals of this paper is that
prescription (3) allows us to perform long evolutions with constraints
converging to zero when the resolution is increased when adopting sixth-order
accurate finite differences. Studying the interplay between dissipation,
convergence order of the numerical scheme, and boundary conditions is left for
future works.

For completeness, in Appendix~\ref{sec:failed-attempts}, we report two
formulations that we experimented with to further improve the stability of the
evolutions but did not result in additional improvements.

\subsection{Contribution of the electromagnetic fields to the spin of a black hole}
\label{sec:contribution-spin-em}

The quasi-isolated horizon formalism is widely used in numerical relativity to
determine quasi-local physical properties (such as mass or angular momentum) of
black holes~\cite{Krishnan2002,Dreyer2003}. However, the majority of previous
simulations focused on the case with gravitational fields only, and hence,
electromagnetic contributions to black hole quasi-local properties have not
been considered. In~\cite{Bozzola2019}, we presented the code
\texttt{QuasiLocalMeasuresEM}, an extended version of
\texttt{QuasiLocalMeasures} that implements the formalism for the full
Einstein-Maxwell theory~\cite{Ashtekar2001}. The main differences with respect
to the purely gravitational case are in the computation of the mass and the
angular momentum of a black hole. The full formalism also provides a quasi-local
way to compute the charge of a black hole. Here, we focus on spin (for a
complete discussion, see Sec.~II~C in~\cite{Bozzola2019}).

\begin{figure}[htbp]
  \centering
  \includegraphics{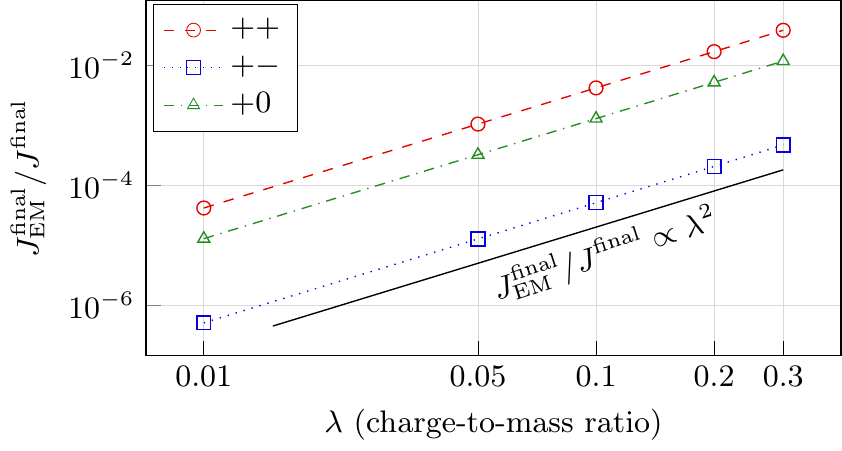}
  \caption{Contribution of the electromagnetic fields to the spin of the final
    black hole. The relative importance of the spin due to electromagnetic
    fields $J_{\text{EM}}$ scales as the charge-to-mass ratio squared. For
    details of the formalism, see Sec.~II~C in~\cite{Bozzola2019}.}
  \label{fig:final_spin_em}
\end{figure}

While the expression for angular momentum at infinity does not depend on
electromagnetic fields~\cite{Ashtekar1978}, the quasi-local formula
does~\cite{Ashtekar2001}. So, to compute the angular momentum $J$ of a charged
black hole, one must calculate two terms: $J_{\text{GR}}$ and $J_{\text{EM}}$
(see, Sec.~II~C in~\cite{Bozzola2019} for more details). The computation of
$J_{\text{GR}}$ is implemented in the \texttt{QuasiLocalMeasures} thorn of the
\texttt{Einstein Toolkit}. \texttt{QuasiLocalMeasuresEM} computes
$J_{\text{EM}}$ as well. We find that this contribution can be a significant
fraction on the total spin. In Fig.~\ref{fig:final_spin_em}, we show the ratio
between $J_{\text{EM}}$ and the total angular momentum of the remnant black hole
forming in our binary simulations
$J^{\text{final}}_{\text{EM}}\slash J^{\text{final}}$ for the various simulations in
our set. The maximum value in our simulations is with $\lambda^{+}_{+} = \num{0.3}$,
where we find that
$J^{\text{final}}_{\text{EM}}\slash J^{\text{final}} \approx \SI{3.8}{\percent}$.
Fig.~\ref{fig:final_spin_em} shows that this ratio depends quadratically on the
charge-to-mass ratio of the final black hole. A fit confirms quantitatively that
this is the case.

\begin{figure}[htbp]
  \centering
  \includegraphics{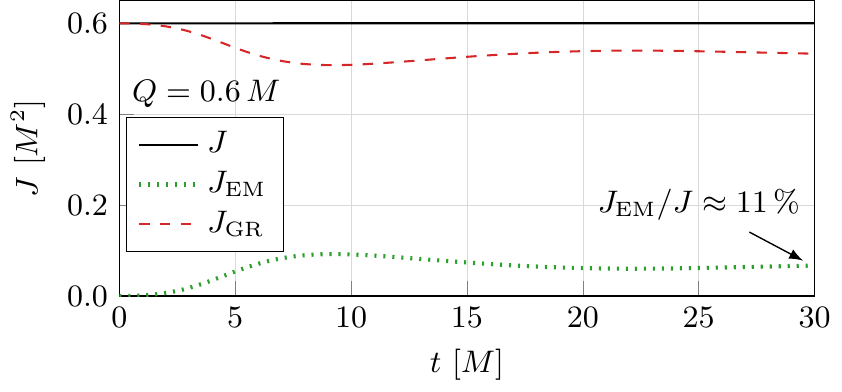}
  \caption{Various contributions to the angular momentum of a rapidly spinning
    and highly charged black hole with $J=\SI{0.6}{\admmass\squared}$ and
    $Q=\SI{0.6}{\admmass}$. As a result of the choice of the initial
    electromagnetic fields, \texttt{TwoChargedPunctures}~\cite{Bozzola2019}
    generates initial data with $J_{\text{EM}} = 0$. During the evolution, the
    initial data relax to a non-zero $J_{\text{EM}}$ while $J$ remains
    conserved.}
  \label{fig:spin_em_evolution}
\end{figure}

\texttt{TwoChargedPunctures} always generates (binary) black hole initial data
with $J_{\text{EM}} = 0$, because of the choice of the initial electromagnetic
fields~\cite{Bozzola2019}. Soon after the evolution starts, the initial data
relax to a non-zero $J_{\text{EM}}$ on a few light-crossing times while keeping
the total angular momentum constant.  In Fig.~\ref{fig:spin_em_evolution}, we
show how the two contributions to the spin behave for the case of a highly
charged, rapidly spinning black hole with $Q=\SI{0.6}{\admmass}$ and
$J=\SI{0.6}{\admmass\squared}$. In this case, after the initial data relax
$J^{\text{final}}_{\text{EM}}\slash J^{\text{final}}$ is approximately
\SI{11}{\percent}. Monitoring this quantity also provides a way to quantify when
the initial data has relaxed.

\subsection{Wave extraction}
\label{sec:grav-electr-waves}

We adopt the Newman-Penrose formalism~\cite{Newman1962} with the same
conventions as in~\cite{Zilhao2015} to extract gravitational and electromagnetic
waves from our simulations. In this subsection, we review the formalism with
focus on the electromagnetic sector. The Newman-Penrose formalism of the
gravitational sector is a standard topic in the literature (e.g.~\cite{Ruiz2008,
  Bishop2016,Shibata2016b}). Details on the numerical implementation of the
calculation of the various quantities presented below can be found
in~\cite{Zilhao2015} and~\cite{Witek2010}.

We consider a null tetrad of complex vectors $k^a, l^a, m^a$ and $ {\conj{m}}^a$
with null mutual inner products except for $- k^a l_a = 1 = m^a \conj{m}_a $. We
also define the Newman-Penrose scalars constructed with the Faraday
tensor~\cite{Teukolsky1973}
\begin{subequations}
  \label{eq:NP-scalars-EM}
  \begin{align}
    \Phi_0 & = F_{ab} l^a m^b \,, \\
    \Phi_1 & = \frac{1}{2} F_{ab} (l^a k^b + {\conj{m}}^a m^b ) \,, \\
    \Phi_2 & = F_{ab} {\conj{m}}^a k^b \,.
  \end{align}
\end{subequations}
Appendix~\ref{sec:electr-newm-penr} connects $\Phi_{0}$, $\Phi_{1}$ and
$\Phi_{2}$ with the electric and magnetic fields in flat spacetime and provides
physical arguments to build intuition on what the different scalars represent.

In asymptotically flat spacetimes, the Newman-Penrose scalars have a known
fall-off behavior at large spatial distances (the so-called ``peeling
theorem''~\cite{Newman1968})
\begin{equation}
  \label{eq:NP-scalars-EM-falloff}
  \Phi_0  \sim \frac{1}{r^3} \,, \qquad \Phi_1  \sim \frac{1}{r^2} \,, \qquad     \Phi_2  \sim \frac{1}{r} \,,
\end{equation}
where the $\sim$ indicates the asymptotic behavior as $r\to\infty$. We will use these
properties to find which terms are important for computations at infinity and
which are not.

By use of Eqs.~\eqref{eq:NP-scalars-EM}, we can express the electromagnetic field
tensor
as
\begin{align}
  F_{ab} &= 2  \left[  \Phi_1 (k_{[a}l_{b]} + m_{[a}\conj{m}_{b]}) \,  +   %
  \Phi_2  l_{[a}m_{b]} + \Phi_0 \conj{m}_{[a} k_{b]} \right]  \nonumber\\  %
  & \quad + \text{complex conjugate}\,,
\end{align}
where ``complex conjugate'' refers to the term in square brackets, and square
brackets next to indices imply antisymmetrization. Considering the
electromagnetic stress-energy tensor
\begin{equation}
  \label{eq:Tmunu-EM}
  4 \pi T_{ab}^{\mathrm{EM}} =  F_{ac} F_{bd} g^{cd} -
  \frac{1}{4}g_{ab} F_{cd} F^{cd}\,,
\end{equation}
we can express this in terms of the Newman-Penrose scalars~\cite{Teukolsky1973}
\begin{align}
  \label{eq:Tmunu-NP}
  4 \pi T_{ab}^{\mathrm{EM}} & =  \biggr[ \Phi_0 \conj{\Phi}_0 k_a k_b +
  2 \Phi_1 \conj{\Phi}_1 \left( l_{(a} k_{b)} +   m_{(a} \conj{m}_{b)}  \right)  \nonumber \\  %
  &   \quad~~ + \Phi_2 \conj{\Phi}_2 l_a l_b -   4 \conj{\Phi}_0 \Phi_1 k_{(a} m_{b)}            \\  %
  &   \quad~~ - 4 \conj{\Phi}_1 \Phi_2 l_{(a} m_{b)} +2 \Phi_2 \conj{\Phi}_0 m_{a} m_{b} \biggr] \nonumber \\  %
  & \quad +  \text{complex conjugate} \,, \nonumber
\end{align}
where the complex conjugate is of the term in square brackets.

Assuming asymptotically spherical coordinates $(r,\theta,\varphi)$ centered on the
Cartesian grid and oriented along the $z$ direction and $t$ along the normal to
the hypersurfaces, we can choose as null tetrad
\begin{subequations}
    \label{eq:null-tetrad}
  \begin{align}
    k^a          & = \frac{1}{\sqrt{2}} (e^a_{\hat{t}} - e^a_{\hat{r}}) \,,  \\
    l^a          & = \frac{1}{\sqrt{2}} (e^a_{\hat{t}} + e^a_{\hat{r}}) \,,  \\
    m^a          & = \frac{1}{\sqrt{2}} (e^a_{\hat{\theta}} + ie^a_{\hat{\varphi}}) \,, \\
    {\conj{m}}^a & = \frac{1}{\sqrt{2}} (e^a_{\hat{\theta}} - ie^a_{\hat{\varphi}}) \,,
  \end{align}
\end{subequations}
with
$\vec{e}_{\hat{t}}, \vec{e}_{\hat{r}}, \vec{e}_{\hat{\theta}}, \vec{e}_{\hat{\phi}}$
orthonormal non-coordinate basis. Note that in the coordinate basis, it holds
that $m_a \sim r$.

The energy and angular momentum fluxes per solid angle carried away by outgoing
electromagnetic waves at infinity is given by
\begin{subequations}
    \label{eq:r2Tmunu-inf}
  \begin{align}
    \frac{\d{}^2 E}{\d t \d \Omega} &= \lim_{r\to+\infty} r^2 T^{r}_{\;\;t}\,,  \label{eq:r2Tmunu-inf-E}\\
    \frac{\d{}^2 L^z}{\d t \d \Omega} &= \lim_{r\to+\infty} r^2 T^{r}_{\;\;\varphi}\,.
  \end{align}
\end{subequations}
Here we focus on the $z$ component of the angular momentum. These quantities can
be expressed in terms of the Newman-Penrose scalars by use of
Eqs.~\eqref{eq:NP-scalars-EM-falloff} and~\eqref{eq:Tmunu-NP}. The only non-zero
contribution to the energy flux arises from the term
$\abs {\Phi_2}^2 l_t l_r \slash 2 \pi$. Hence,
\begin{equation}
  \label{eq:energy-em}
  \frac{\d{}^2 E}{\d t \d \Omega} = \lim_{r\to+\infty} r^2 T^r_{\;\;t} =\lim_{r\to+\infty}
  \frac{r^2}{2\pi} \abs {\Phi_2}^2 \,.
\end{equation}
Considering the fall-off behavior of the Newman-Penrose scalars in Eqs.~\eqref{eq:NP-scalars-EM-falloff}, the angular momentum flux becomes
\begin{equation}
  \label{eq:non-zero-Lz}
  \lim_{r\to+\infty} r^2 T^{r}_{\;\;\varphi} = \lim_{r\to+\infty} -\frac{r^2}{2\pi} \left( \conj{\Phi}_1 \Phi_2 l_r m_\varphi + \Phi_1 \conj{\Phi}_2 l_r \conj{m}_\varphi   \right)\,,
\end{equation}
which can be rewritten as
\begin{align*}
  \label{eq:non-zero-Lz-second-step}
  \lim_{r\to+\infty} r^2 T^{r}_{\;\;\varphi} & =\lim_{r\to+\infty} -\frac{r^2}{4\pi} \left( \conj{\Phi}_1 \Phi_2 i r \sin \theta - \Phi_1 \conj{\Phi}_2 i r \sin \theta  \right) \\
                                & =\lim_{r\to+\infty} -\frac{ir^3\sin \theta}{4 \pi} \left( \conj{\Phi}_1 \Phi_2 - \Phi_1 \conj{\Phi}_2  \right)               \\
                                & =\lim_{r\to+\infty} \frac{r^3\sin \theta \Im[\conj{\Phi}_1 \Phi_2]}{2\pi}\,,
\end{align*}
where $\Im[z]$ is the imaginary part of the complex number $z$. Therefore, the
flux of angular momentum that crosses a sphere at large radius $r$ is
\begin{equation}
  \label{eq:flux-L}
  \frac{\d{}^2 L^z}{\d t \d \Omega} =\frac{1}{2\pi} r^3\sin \theta \Im[\Phi_1 \conj{\Phi}_2] \,.
\end{equation}
Here we reach the same conclusion as in~\cite{Ashtekar2017}: the flux of
angular momentum does not depend only on the radiative degrees of freedom
(encoded in $\Phi_2$), but there is also a Coulombic contribution (encoded in
$\Phi_1$)~\cite{Bonga2018}. This is a striking difference compared to gravitational
waves, for which the information contained in the radiative degrees of freedom,
i.e.\ $\Psi_4$, is sufficient for computing the flux of angular
momentum~\cite{Bonga2019}.

We perform a decomposition in spin-weighted spherical harmonics $Y_{\ell m}^s$ with
spin $s = -2, -1, 0$
\begin{subequations}
  \label{eq:spin-weighted-spherical-harmonics}
  \begin{align}
    \Phi_1(r,\theta,\varphi) &= \sum_{\ell\ge0} \sum_{-\ell\le m \le \ell} \phi^{\ell m}_1 Y_{\ell m}^0 \,, \\
    \Phi_2(r,\theta,\varphi) &= \sum_{\ell\ge1} \sum_{-\ell\le m \le \ell} \phi^{\ell m}_2 Y_{\ell m}^{-1} \,, \\
    \Psi_4(r,\theta,\varphi) &= \sum_{\ell\ge2} \sum_{-\ell\le m \le \ell} \psi^{\ell m}_4 Y_{\ell m}^{-2} \,.
  \end{align}
\end{subequations}
Using the orthogonality relations between spin-weighted spherical harmonics with
the same spin, we can write~\cite{Baumgarte:2010nu}
\begin{subequations}
\label{eq:energy-momentum-spin-weighted}
  \begin{align}
    \frac{\d{} E_{\text{EM}}}{\d t} &= \lim_{r\to+\infty}  \frac{r^2}{4\pi} \sum_{\ell,m} \abs {\phi_2^{\ell m}}^2 \,, \\
    \frac{\d{} E_{\text{GW}}}{\d t} &= \lim_{r\to+\infty}  \frac{r^2}{16\pi} \sum_{\ell,m}  \biggl\lvert {\int_{-\infty}^{t} \d t' \psi_4^{\ell m}} \biggr\rvert^{2}  \,, \\
    \frac{\d{} L_{\text{GW}}}{\d t} &= \lim_{r\to+\infty}  \frac{r^2}{16\pi} \sum_{\ell,m} m \Im \Biggl[ \left({\int_{-\infty}^{t} \d t' \psi_4^{\ell m}}\right) \\
    &\quad \times \left({\int_{-\infty}^{t} \d t' \int_{-\infty}^{t'} \d t'' \psi_4^{\ell,m*}}\right)  \Biggr] \,. \nonumber
  \end{align}
\end{subequations}
The sums on $m$ go from $-l$ to $l$. In practice, we truncate the expansion to
$l = 8$ and we use a finite extraction radius. There is no simple relation
between the angular momentum lost by electromagnetic waves and the multipolar
components $\phi_2^{\ell m}$ and $\phi_1^{\ell m}$ when decomposed with the respective
spin-weight, so we implemented directly Eq.~\eqref{eq:flux-L}. We tested this
new diagnostic for the flux of angular momentum against the analytical Michel
solution~\cite{Michel1973,Gralla2014} (see, Appendix~\ref{sec:michel-solution}).

\section{Quadrupole Approximation Model}
\label{sec:newtonian-model}

The simplest waveform model for charged binary black holes is obtained from the
Keplerian motion of charged massive particles in Newtonian physics with the
inclusion of radiation reaction. We indicate this model with the initials QA
(\emph{quadrupole approximation}). We will also refer to it as the ``Newtonian
model''.

Let us consider two point particles with masses $m_{1}$, $m_{2}$ and charges
$q_{1}$, $q_{2}$ on a circular orbit at separation $d$. The total energy
(kinetic + gravitational + electrostatic) of the system $E$ is
\begin{equation}
  \label{eq:energy}
  E = -\frac{m_{1}m_{2}}{2d} -\frac{q_{1}q_{2}}{2d} = -(1 - \lambda_{1}\lambda_{2})\frac{m_{1}m_{2}}{2d}\,,
\end{equation}
where $\lambda_{i} = {q_{i}}\slash{m_{i}}$ is the charge-to-mass ratio of the
$i$-th particle.

The system loses energy by emission of gravitational and electromagnetic waves.
In this work, we will restrict to circular orbits and only consider dipole and
quadrupole electromagnetic waves, and quadrupole gravitational
waves~\cite{Liu2020,Christiansen2020}
\begin{subequations}
    \label{eq:loss}
  \begin{align}
  \frac{\d E^{\text{dip}}_{\text{EM}}}{\d t} &= \frac{2}{3} {(\lambda_{1} - \lambda_{2})}^{2}{(1 - \lambda_{1} \lambda_{2})}^{2} \frac{m_{1}^{2} m_{2}^{2}}{d^{4}}\,, \label{eq:dedt1} \\
  \frac{\d E^{\text{quad}}_{\text{GW}}}{\d t} &= \frac{32}{5} {(1 - \lambda_{1} \lambda_{2})}^{3} \frac{m_{1}^{2} m_{2}^{2} M}{d^{5}}\,,  \label{eq:dedt2}\\
  \frac{\d E^{\text{quad}}_{\text{EM}}}{\d t} &= {\left(\frac{m_{2} \lambda_{1}}{2 M} + \frac{m_{1} \lambda_{2}}{2 M} \right)}^{2}\frac{\d E^{\text{quad}}_{\text{GW}}}{\d t}\,.  \label{eq:dedt3}
\end{align}
\end{subequations}
This can be extended to eccentric orbits~\cite{Liu2020}. The angular momentum
carried away by gravitational and electromagnetic waves in the dipole and
quadrupolar channels is given by~\cite{Liu2020}
\begin{subequations}
  \label{eq:lossang}
  \begin{align}
  \frac{\d J^{\text{dip}}_{\text{EM}}}{\d t} &= \frac{\d E^{\text{dip}}_{\text{EM}}}{\d t} \biggr\slash \sqrt{(1 - \lambda_{1} \lambda_{2})\frac{M}{d^3}}\,, \label{eq:Jem}\\
  \frac{\d J^{\text{quad}}_{\text{GW}}}{\d t} &= \frac{\d E^{\text{quad}}_{\text{GW}}}{\d t} \biggr\slash \sqrt{(1 - \lambda_{1} \lambda_{2})\frac{M}{d^3}}\,.
\end{align}
\end{subequations}
Note that the denominator is the orbital angular velocity of the binary.

We can derive the equation of motion by taking the derivative of
Eq.~\eqref{eq:energy} and applying the chain rule
\begin{equation}
  \label{eq:eom0}
  \frac{\d E}{\d t} = (1 - \lambda_{1}\lambda_{2})\frac{m_{1}m_{2}}{2d^{2}} \frac{\d d}{\d t}\,.
\end{equation}
Therefore,
\begin{equation}
  \label{eq:eom}
  \frac{\d d}{\d t} = \frac{2 d^{2}}{(1 - \lambda_{1}\lambda_{2})m_{1}m_{2}} \frac{\d E}{\d t}\,.
\end{equation}
To respect energy conservation, ${\d E}\slash{\d t}$ is given by the total energy
loss via electromagnetic and gravitational radiation provided by
Eqs.~\eqref{eq:dedt1},~\eqref{eq:dedt2},~\eqref{eq:dedt3}. The resulting
equation of motion~\eqref{eq:eom} cannot be solved analytically in closed form
for a non-zero charge. Here, we solve it numerically with the LSODA
solver~\cite{LSODA} of ODEPACK~\cite{hindmarsh1982odepack} through the SciPy
interface~\cite{SciPy}. The time integration is performed with a timestep that
is proportional to the orbital separation. We continue the integration up to
$d = \SI{5}{\admmass}$, which is an average radius of the Innermost Stable
Circular Orbit (ISCO) for neutral particles around Kerr-Newman black holes (more
on this in Sec.~\ref{sec:spin}).

The solution of Eq.~\eqref{eq:eom} provides the time evolution of the orbital
separation. Then, using Eqs.~\eqref{eq:loss},~\eqref{eq:lossang}, we can compute
the energy and angular momentum lost by gravitational and electromagnetic
waves. We can also compute the gravitational-wave strain in the quadrupole
approximation~\cite{Maggiore2007, Wang2020} measured at distance $r$ from the
binary
\begin{subequations}
  \begin{align}
    r h_{+}&=4 {(1 - \lambda_{1} \lambda_{2})}^{\frac{5}{3}} \frac{m_{1} m_{2}}{M^{\frac{1}{3}}} {(\pi f_{\text{GW}})}^{\frac{2}{3}} \cos \phi_{\text{GW}}\,, \\
    r h_{\times}&= 4 {(1 - \lambda_{1} \lambda_{2})}^{\frac{5}{3}} \frac{m_{1} m_{2}}{M^{\frac{1}{3}}} {(\pi f_{\text{GW}})}^{\frac{2}{3}} \sin \phi_{\text{GW}}\,.
  \end{align}
\end{subequations}
The frequency of the gravitational waves $f_{\text{GW}}$ is twice the orbital frequency,
\begin{equation}
  \label{eq:f-gw}
  f_{\text{GW}}(d) = \frac{1}{\pi} \sqrt{\frac{(1 - \lambda_{1}\lambda_{2})M}{d^{3}}}\,.
\end{equation}
Defining $\omega_{\text{GW}}(d) = 2 \pi f_{\text{GW}}(d)$, the phase of the
gravitational waves $\phi_{\text{GW}}$ is
\begin{equation}
  \label{eq:phi-gw}
  \phi_{\text{GW}} = \phi_{0} + \int_{d_0}^{d} \frac{\omega_{\text{GW}}(R)}{\dot{d}(R)} \d R \,,
\end{equation}
where $\phi_{0}$ is an arbitrary initial phase and $\dot{d}(R)$ is the time
derivative of $d$ evaluated at orbital separation $R$, which can be computed
from Eq.~\eqref{eq:eom}.

Finally, we note that the computational requirements of the QA model are
negligible compared to numerical-relativity simulations (as Eq.~\eqref{eq:eom}
is a single ordinary differential equation). In certain limits, the equations
can be even solved analytically~\cite{Christiansen2020}.

\section{The remnant black hole}
\label{sec:final-black-hole}

The properties of the remnant black hole that forms following a binary black
hole merger are key to testing general
relativity~\cite{Berti2006,Yang:2017zxs,Berti2018}. Perturbation theory applies
to the post-merger black hole, and it is well established that perturbed black
holes settle by undergoing characteristic damped oscillations. This is known as
quasi-normal-mode ringing. In general relativity, the complex frequency of these
oscillations is completely determined by the black hole mass, spin, and charge.

In this section, we first present the details of an existing method to estimate
the spin of the remnant black hole forming following mergers of charged black
holes. Next, we discuss quasi-normal-modes. The two discussions are closely
related as the quasi-normal-modes depend on the properties of the black
hole.\footnote{Another way in which the two discussions are linked is that the
  study of geodesic motion can be used to estimate the quasi-normal-mode
  frequencies from the light-ring
  properties~\cite{Berti2005,Cardoso2009,Khanna2017}.}

For the values of charge-to-mass ratios $\lambda$ considered in this work, the
quasi-normal-modes are primarily determined by the spin of the final black hole,
because $\lambda$ is not large enough to matter. However, the final spin depends
on the charge of the binary components. Even when the total charge is zero,
e.g.~the two black holes have equal and opposite charges, the final spin is
expected to be different from the uncharged case, allowing, in principle, to
distinguish the two scenarios.

\subsection{Remnant black hole spin model}
\label{sec:spin}

A simple way to estimate the spin of the remnant black hole forming following
the merger of two black holes on a quasi-circular orbit is to invoke
conservation arguments (this method is sometimes known with the acronym
BKL~\cite{Buonanno2008}, from the names of the authors).  The approach aims to
be \SI{10}{\percent} accurate~\cite{Buonanno2008}, and while it was first
applied to Kerr-Newman black holes in~\cite{Jaiakson2017}, the performance of
the approximation has not been tested in the case of charged black hole
inspirals. Therefore, it is unknown if the accuracy goal is met for more generic
cases. Here we use our quasi-circular inspiral calculations to gauge the
accuracy of the BKL approach.

The basic assumption of the method is that during an inspiral the black-hole
orbital separation shrinks due emission of gravitational waves until the ISCO is
reached, at which point the two black holes plunge. During the plunge, little
angular momentum is lost, so one can estimate the spin of the remnant black hole
by studying the ISCO.\@

Consider the merger of two non-spinning black holes with mass $m_{1}$, $m_{2}$
and charge $q_{1}$, $q_{2}$. The total mass and charge of the system are
\begin{equation}
  \label{eq:total-m-q}
  M = m_{1} + m_{2}\,, \quad Q = q_{1} + q_{2}\,.
\end{equation}
Given that the energy lost via gravitational waves is of order of a few percent
of $M$ (smaller than the target accuracy of \SI{10}{\percent}), the BKL method assumes
that the total mass is conserved. Therefore, the final black hole has mass $M$.
Attempts to include energy loss in the uncharged case were
made~\cite{Kesden2008}, but we will not consider this here. Charge is exactly
conserved, so the final black hole must have charge $Q$. To determine the
remnant black hole spin parameter $a$, one invokes angular momentum
conservation. The BKL model postulates that the final angular momentum $Ma$ is
exactly the same as the orbital angular momentum at the onset of the plunge.
This quantity is then estimated by considering the motion in the spacetime of
the remnant black hole of a test particle with mass $\mu$ and charge $q$ given by
\begin{equation}
  \label{eq:reduced-mass-charge}
  \mu = \frac{m_{1} m_{2}}{M}\,, \quad q = \frac{q_{1} q_{2}}{Q}\,,
\end{equation}
where $\mu$ and $q$ are the reduced mass and charge of the binary system
(namely, the mass and charge of the equivalent effective one-body problem).

Let $L_{M, a, Q}$ be the ISCO angular momentum of a test particle with mass
$\mu$ and charge $q$ in a Kerr-Newman spacetime with mass $M$, spin $a$, and
charge $Q$, then the BKL approach sets
\begin{equation}
  \label{eq:convervation-angular-momentum}
  M a = L_{M, a, Q}(r_{\text{ISCO}})\,.
\end{equation}
If $l=L_{M, a, Q}(r_{\text{ISCO}})/\mu$ is the test particle specific angular
momentum, then we have
\begin{equation}
  \label{eq:convervation-angular-momentum-a}
  a = \nu l_{M, a, Q}(r_{\text{ISCO}}) \,,
\end{equation}
where $\nu = \mu \slash M = (m_{1} m_{2})\slash M^{2}$ is the symmetric mass
ratio.  Equation~\eqref{eq:convervation-angular-momentum-a} determines the spin
$a$ of the final black hole. It is straightforward to extend the method to
consider spinning binaries by adding the contribution of the individual spins to
the total angular momentum in Eq.~\eqref{eq:convervation-angular-momentum}, but
we will not do this here, because our numerical relativity simulations of
charged binaries do not involve spin.

To solve Eq.~\eqref{eq:convervation-angular-momentum-a}, we need to compute
the specific angular momentum of the test particle with charge $q$ in a circular
orbit at radius $r_{\text{ISCO}}$. We do this in the standard way by defining an
effective radial potential $V_{\text{eff}}(r)$.

In Boyer-Lindquist coordinates $(t,r,\theta,\phi)$, the line element $\d s^2$ of
a Kerr-Newman black hole with mass $M$, spin $a$, and charge $Q$ is given
by~\cite{Kerr1963,Newman1965}
\begin{equation}
  \label{eq:Kerr-Newman-ds}
  \begin{split}
    \d s^2 =& -\frac{\Delta - a^2 \sin^2 \theta}{\rho^2} \d t^2  + \frac{\rho^2}{\Delta} \d
    r^2 +  \rho^2 \d \theta^2 \\
    &- 2 a \sin^2 \theta \frac{(r^2 + a^2
      -\Delta)}{\rho^2} \d t \d \phi                         \\
    &+ \frac{{(r^2 + a^2)}^2 - \Delta a^2 \sin^2 \theta}{\rho^2} \sin^2 \theta \d \phi^2\,,
  \end{split}
\end{equation}
with
\begin{subequations}
  \label{eq:Kerr-Newman-symbols}
  \begin{align}
    \rho^2 & = r^2 + a^2 \cos^2 \theta\,, \\
    \Delta   & = r^2 - 2Mr + a^2 + Q^2\,.
  \end{align}
\end{subequations}
The electromagnetic vector potential is
\begin{equation}
  \label{eq:Ker-Newman-em}
  A = - \frac{Qr}{\rho^2}(\d t - a \sin^2 \theta \d \phi)\,.
\end{equation}
For Kerr-Newman black holes, $V_{\text{eff}}$ is given
by~\cite{Jaiakson2017,Siahaan2020}
\begin{equation}
  \label{eq:Veff}
  V_{\text{eff}} = \frac{g_{tt} {\tilde{l}}^{2} + 2 g_{t\phi}\tilde{l} \tilde{\varepsilon} + g_{\phi\phi}{\tilde{\varepsilon}}^{2} - \Delta}{g_{rr} \Delta}\,,
\end{equation}
with
\begin{subequations}
\begin{align}
  \tilde{l} & = l + q A_{\phi}\,, \\
  \tilde{\varepsilon} & = \varepsilon - q A_{t}\,,
\end{align}
\end{subequations}
where $\varepsilon$ and $l$ are the specific energy and specific angular
momentum, respectively. The properties of the ISCO are found solving the
following equations simultaneously
\begin{subequations}
\begin{align}
  \label{eq:isco}
  V_{\text{eff}}(r_{\text{ISCO}}) &= 0\,, \\
  \frac{\d V_{\text{eff}}}{\d r}(r_{\text{ISCO}}) &= 0\,, \\
  \frac{\d{}^{2} V_{\text{eff}}}{\d r^{2}}(r_{\text{ISCO}}) &= 0\,.
\end{align}
\end{subequations}
for $r_{\text{ISCO}}$, $ \varepsilon(r_{\text{ISCO}})$ and $l(r_{\text{ISCO}})$.

For the estimate of the spin, we are particularly interested in
$l(r_{\text{ISCO}})$.  Once that is known, we use a root-finding method to solve
numerically Eq.~\eqref{eq:convervation-angular-momentum} and find the spin
of the final black hole.

\subsection{Quasi-normal-modes}
\label{sec:quasi-normal-modes}

Following merger, the remnant black hole settles by undergoing quasi-normal mode
ringing, i.e., damped oscillations with specific frequencies $\omega$ and decay
times $\tau$~\cite{Berti2009}.\footnote{Here we are considering relatively small
  charge, so we will only be focusing on the gravitational quasi-normal-modes.}
During the ringdown phase, the Newman-Penrose scalar $\Psi_{4}$ looks like
\begin{equation}
  \label{eq:qnm}
  \Psi_{4}(t,r) \sim \sum_{\ell mn} A_{\ell m}(r) \mathrm{e}^{-t\slash \tau_{\ell mn}} \sin (\omega_{\ell mn} t)\,,
\end{equation}
with $l$, $m$ being the multipolar mode numbers and $n$ the overtone number; $A,
\tau$, and $\omega$ are the characteristic amplitude, the decay time, and the
frequencies of the quasi-normal modes. These values depend on the mass, spin,
and charge of the black hole in a known way~\cite{Berti2006,Berti2018}.  In this
work, we are interested in exploring the charge information contained in the
ringdown waveforms. In particular, we test whether it is possible to tell
whether a merging binary had charge by looking at the post-merger signal alone.

For Schwarzschild and Kerr black holes, the values of $\omega_{\ell mn}$ and
$\tau_{\ell mn}$ are tabulated~\cite{bertiqnm:web,cardosoqnm:web} or available
in public codes, like the one we use here--\texttt{qnm}~\cite{Stein:2019mop}.
For generic Kerr-Newman solutions, while the problem has been
solved~\cite{Dias2015}, such tables are not publicly available. However, since
the simulations in our set have relatively small charge-to-mass ratio, we can
work in the small charge limit and we can use the equations provided
in~\cite{Cardoso2016b} for the $\ell=2$, $m=2$, $n=0$ quasi-normal mode (which
typically dominates~\cite{Berti2007, Barausse2012}). As
shown in~\cite{Cardoso2016b}, for a Kerr-Newman black hole with charge-to-mass
ratio $\lambda$ and dimensionless spin $\chi$, the first correction to
$\omega_{220}$ and $\tau_{220}$ with respect to the uncharged values
$\omega_{220}^{\lambda=0}$ and $\tau_{220}^{\lambda=0}$ is:
\begin{subequations}
  \label{eq:qnm-kn}
\begin{align}
  \frac{\delta \omega_{220}}{\omega_{220}^{\lambda=0}} &= \lambda^{2} \left[-\num{0.2812}-\num{0.0243}\chi + \frac{\num{0.3506}}{{(1 - \chi)}^{\num{0.505}}} \right]\,,\\
  \frac{\delta \tau_{220}}{\tau_{220}^{\lambda=0}} &= - \lambda^{2} \left[\num{0.1075}+\num{0.08923}\chi + \num{0.02314}\chi^{2}  \vphantom{\frac{\num{0.07585}}{{(1 - \chi)}^{\num{0.505}}}} \right.\\
                             & \quad\quad\quad~ \left. + \num{0.09443}\chi^{3} - \frac{\num{0.07585}}{{(1 - \chi)}^{\num{1.2716}}}  \right]\,, \nonumber
\end{align}
\end{subequations}
with $\delta \omega_{220} = \omega_{220} - \omega_{220}^{\lambda=0}$ and
$\delta \tau_{220} = \tau_{220} - \tau_{220}^{\lambda=0}$. In this work, we use
\texttt{qnm}~\cite{Stein:2019mop} to compute $\omega_{220}^{\lambda=0}$ and
$\tau_{220}^{\lambda=0}$ and Eqs.~\eqref{eq:qnm-kn} to compute the quasi-normal
modes for our charged remnants.

We can plug a representative remnant black hole spin value $\chi=0.67$ in
Eq.~\eqref{eq:qnm-kn} to gain some insight on the effect of charge, which yields
${\delta \omega_{220}}\slash{\omega_{220}^{\lambda=0}} \approx \lambda^{2}
\slash 3$, and ${\delta \tau_{220}}\slash{\tau_{220}^{\lambda=0}} \approx
\lambda^{2} \slash 10$. For the values of $\lambda$ treated in our simulations,
if we assumed that all black holes had the same final mass and spin, then the
deviations from the Kerr quasi-normal modes are at most at the percent level. The
deviations are maximized for larger $\lambda$, so one would expect that the
simulation with $\lambda^{+}_{+}$ is the easiest to constrain. However, if one
wanted to tackle the question ``Can we tell from the ringdown if the binary was
charged?'' the problem is more complicated and one needs to consider the
interplay between mass, spin, and charge. We discuss this in
Sec.~\ref{sec:quasi-normal-modes-1}.

\section{Simulation Results and Discussion}
\label{sec:results}

In this section, we describe the results from our black hole binary simulations
through the inspiral, merger and ringdown phases. We start by analyzing the
inspiral (Sec.~\ref{sec:inspiral}). We compare the non-linear solutions with the
QA model, finding that the Newtonian approach always overestimates observable
quantities by \num{20}--\SI{100}{\percent}. Next, we present our
numerical-relativity simulations up to merger (Sec.~\ref{sec:up-merger}),
discussing properties of the emission (Sec.~\ref{sec:properties-emission}) and
detectability of charge by future gravitational wave observatories
(Sec.~\ref{sec:detect-charge-with}). Finally, we explore the ringdown phase
(Sec.~\ref{sec:prop-final-black}), testing the BKL approach
(Sec.~\ref{sec:final-spin}) and discussing quasi-normal-modes
(Sec.~\ref{sec:quasi-normal-modes-1}).

\subsection{Inspiral}
\label{sec:inspiral}

Here, we focus on the inspiral part of our calculations and we assess
the performance of the Newtonian model within the quadrupole approximation
(QA). Since high-order PN and/or effective-one-body waveforms including
electromagnetic fields are not yet available, this model is the most widely used
to study inspirals of charged black holes
(e.g.~\cite{Cardoso2016b,Wang2020,Christiansen2020,Liu2020,Liu2020b,Liu2020c,Cardoso2020Erratum}).

To compare numerical-relativity simulations with the Newtonian model, we
consider a set of quantities of interest. To keep our results gauge-independent,
we analyze these quantities between reference gravitational-wave frequencies
$f_{0}, f_{1}, $ and $f_{2}$, which we choose as follows: $\si{\admmass} f_{0} =
\num{9.61e-3}$, $\si{\admmass} f_{1} = \num{1.76e-2}$, and $\si{\admmass} f_{3}
= \num{3.84e-2}$, where $\si{\admmass}$ is the detector-frame ADM mass. The
frequency $\si{\admmass} f_{0}$ corresponds to approximately \SI{28}{\Hz} for a
binary with source-frame ADM mass \SI{65}{\sunmass} (detector-frame mass of
approximately \SI{70.6}{\sunmass}).  These three frequencies are motivated by
the LIGO sensitivity band with $f_{0}$ corresponding to the onset of the latest
stage of the inspiral, $f_{1}$ to the intermediate phase prior to plunge, and
$f_{2}$ to the plunge. Previous studies used the Newtonian model for LIGO-Virgo
mergers, so our analysis here gauges how this approximate method performs and
allows us to estimate the level of its accuracy. Since our numerical-relativity
simulations scale with the total mass of the system \si{\admmass}, we can target
both stellar-mass and supermassive black hole binaries. In
Table~\ref{tab:freq-mass} we show what frequencies correspond to our reference
frequencies $f_{0}, f_{1}, f_{2}$ for different choices of \si{\admmass}. The
table shows that this work is relevant to LIGO-Virgo as well as LISA
sources~\cite{Amaro2017}. In Sec.~\ref{sec:detect-charge-with}, we discuss
charge detectability by LISA.\@

\begin{table}[htbp]
  \centering
  \caption{Reference frequencies $M f_{0} = \num{9.61e-3}$, $M f_{1} =
    \num{1.76e-2}$, and $M f_{3} = \num{3.84e-2}$ for different values of the
    detector-frame mass \si{\admmass}. This study targets both LIGO-Virgo and
    LISA sources. The choice of $\si{\admmass} = \SI{70.6}{\sunmass}$ is
    inspired by GW150914~\cite{Abbott2016d}. The evolution from $f_{0}$ to
    $f_{1}$ is still not in the most relativistic part of the inspiral. On the
    other hand, $f_{2}$ is reached in the latest stages of the merger, following
    which the black holes plunge.}
  \label{tab:freq-mass}
  \begin{tabular}[t]{lcccc}
    Freq.\ & $\si{\admmass} = \SI{30}{\sunmass}$ & $\si{\admmass} = \SI{70.6}{\sunmass}$ & $\si{\admmass} = \SI{e4}{\sunmass}$  & $\si{\admmass} = \SI{e7}{\sunmass}$\\
    $f_{0}$   & $\SI{65}{\Hz}$    & \SI{28}{\Hz}                        & \SI{0.195}{\Hz}                 & \SI{0.195}{\milli\Hz} \\
    $f_{1}$   & $\SI{120}{\Hz}$    & \SI{51}{\Hz}                        & \SI{0.357}{\Hz}                 & \SI{0.357}{\milli\Hz} \\
    $f_{2}$   & $\SI{260}{\Hz}$    & \SI{111}{\Hz}                       & \SI{0.780}{\Hz}                 & \SI{0.780}{\milli\Hz} \\
  \end{tabular}
\end{table}

We designate by $t_{f}$ the coordinate time at which the gravitational-wave
frequency is $f$, and using our non-linear simulations we compute the error of
the QA model in gauge invariant quantities within the time intervals
$[t_{f_{0}},t_{f_{1}}]$ and $[t_{f_{0}},t_{f_{2}}]$. The quantities we consider
are: gravitational-wave phase, energy and angular momentum lost through
emission, number of gravitational-wave cycles $N_{\text{GW}}$, and the
signal-to-noise ratio (SNR) between $f_{\text{min}}$ and $f_{\text{max}}$ (which
are of particular interest to estimate charge detectability~\cite{Cardoso2016b}).
The latter two quantities are computed as (see
e.g.,~\cite{Maggiore2007,Moore2015})
\begin{subequations}
  \begin{align}
    \label{eq:N-gw-snr}
    N_{\text{GW}} &= \int_{f_{\text{min}}}^{f_{\text{max}}} \frac{f_{\text{GW}}}{\dot{f}_{\text{GW}}} \d f_{\text{GW}}\,, \\
    \text{SNR}^2 &= 4 \int_{f_{\text{min}}}^{f_{\text{max}}}\frac{\abs{\tilde{h}(f)}^{2}}{S_{n}(f)} \d f\,,
  \label{eq:snr}
  \end{align}
\end{subequations}
with $\tilde{h}$ Fourier transform of the strain and $S_{n}(f)$ the power
spectral noise density of the detector. In the rest of the discussion, we focus
on advanced LIGO at design sensitivity.
\begin{figure}[htbp]
  \centering
  \includegraphics{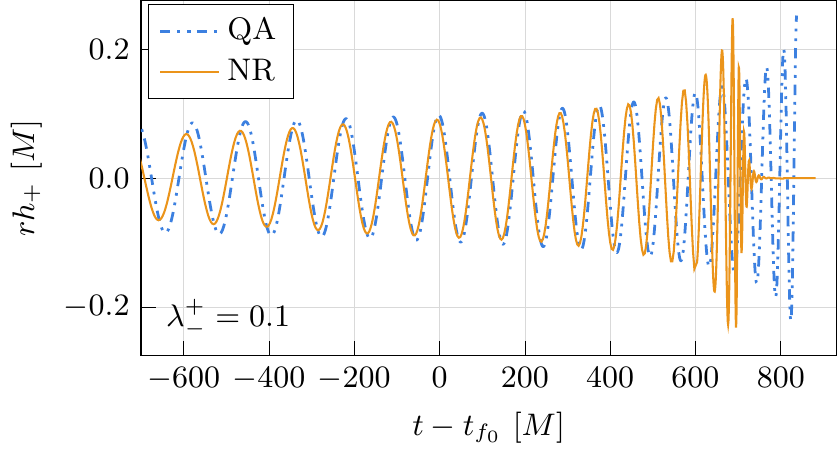}
  \caption{Plus polarization of the gravitational-wave strain produced by
    oppositely charged binary black holes with $\lambda^{+}_{-} = 0.1$ as
    extracted from the Newtonian (QA, blue, dash-dotted line) and full
    numerical-relativity simulation (NR, orange, solid line). The two waveforms
    are aligned a $t_{f_{0}}$, which is when the gravitational wave frequency is
    $f_{0} = \SI{9.61e-3}{\per\admmass}$. The main difference is that the fully
    relativistic simulations predict a faster merger, as they include all
    non-linear terms.}
  \label{fig:gwsignal}
\end{figure}

Before discussing the quantitative differences between the numerical relativity
(NR) and QA models, we first provide a qualitative description.
Figure~\ref{fig:gwsignal} shows the plus polarization of the $\ell=2$, $m=2$ mode
obtained with NR and the Newtonian model for a representative charge black hole
binary with $\lambda^{+}_{-} = 0.1$. The two waveforms shown in the figure are aligned
at $t_{f_{0}}$, when they both have the same gravitational-wave frequency
$f_{0}$. As the plot demonstrates, the QA model provides a decent approximation
to the NR signal up to $t-t_{f_0}\approx\SI{400}{\admmass}$. Shortly after that time,
the black holes merge in the NR simulation. Merger is never captured in the QA
model, and the Newtonian simulation is stopped when the separation is of order
of the ISCO (as in~\cite{Wang2020}). Also, there is substantial de-phasing before
the frequency-alignment time. A major difference between the two models is that
the relativistic simulation predicts a faster merger, because it includes all
non-linear terms. As we will discuss later, this is the fundamental reason why
the QA model overestimates all the interesting physical quantities.

\begin{figure}[htbp]
  \centering
  \includegraphics{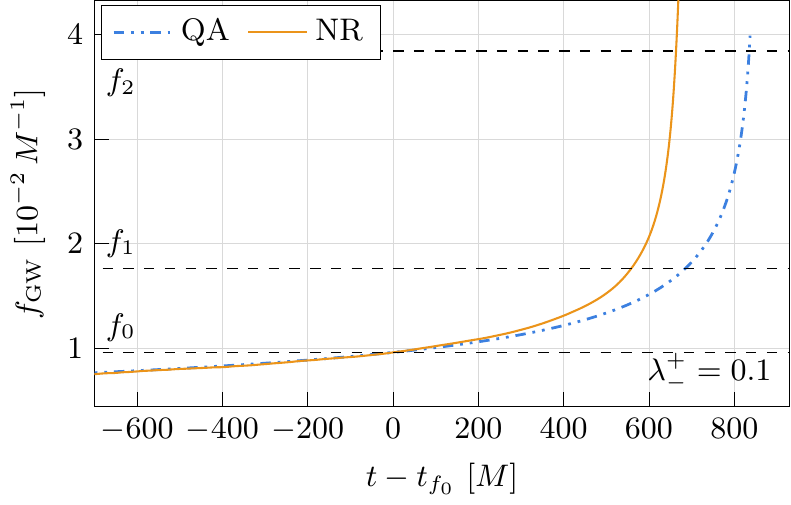}
  \caption{Gravitational-wave frequency evolution for the Newtonian model and
    for the fully general-relativistic simulation. The two simulations are
    aligned at $t_{f_{0}}$, when the gravitational-wave frequency is $f_{0}$.
    Table~\ref{tab:freq-mass} reports the values of $f_{0}, f_{1}$ and
    $f_{2}$.}
  \label{fig:frequency_evolution}
\end{figure}

We further emphasize this point in Fig.~\ref{fig:frequency_evolution}, where we
report the frequency of the gravitational waves in the two models as a function
of time. It is clear that the frequency evolves faster in the non-linear
calculation. As a result, the QA computations spend more time inspiraling, so
this model overestimates all relevant quantities, as it overestimates the time
from $t_{f_{0}}$ to $t_{f_{1}}$ and $t_{f_{2}}$. This is exactly what happens at
the quantitative level, too. Table~\ref{tab:summary} reports the relative error
of the Newtonian calculations with respect to the NR simulations. For each
quantity $\Upsilon$, the error is computed as
\begin{equation}
  \label{eq:rel-error}
  \text{relative error} = \frac{\Upsilon^{\text{QA}} - \Upsilon^{\text{NR}}}{\Upsilon^{\text{NR}}}\,.
\end{equation}
No absolute value is taken: a positive error means that the Newtonian
approximation overestimates $\Upsilon$. The ranges of error reported are across all NR
simulations we performed for this work. As the values in the table demonstrate,
the listed quantities are always overestimated by order \SI{20}{\percent} or more.
However, we note that the QA model always captures the correct order of
magnitude in the amplitude up until a couple of cycles prior to peak
gravitational wave amplitude in the non-linear calculations. Hence, for the
values of $\lambda$ considered here, the model can be used for rough
estimates.\footnote{In~\cite{Shibata2003,Reisswig2011b} it was shown that when
  adopting the quadrupole formula to estimate gravitational waves, but with the
  fluid distribution computed based on general relativistic simulations, the
  errors are of order \SI{20}{\percent} in the amplitude. Here, we show the performance
  of the quadrupole approximation when not coupled to trajectories from
  numerical relativity simulations is significantly worse.}

\begin{center}
\begin{table*}[t]
  \caption{Summary of relative errors for key quantities across all our
    simulations. The relative error for a quantity $\Upsilon$ is computed as
    $(\Upsilon^{\text{QA}} - \Upsilon^{\text{NR}})\slash\Upsilon^{\text{NR}}$, where $\Upsilon^{\text{QA}}$ is
    the quantity in the quadrupole approximation and $\Upsilon^{\text{NR}}$ in the
    simulations. The relative error is with sign: positive error means that the
    Newtonian approximation \emph{overestimates} the quantity. In all cases the
    quadrupole approximation overestimates the quantities. The values of the
    reference frequencies is reported in Table~\ref{tab:freq-mass}. GW and EM
    indicate gravitational and electromagnetic waves respectively. When it comes
    to loss of angular momentum due to electromagnetic waves, the QA model only
    includes the dipolar channel, which is identically zero in the $++$ case.
    Hence, we do not report this error for this quantity.}
  \label{tab:summary}
  \centering
  \begin{tabular}[t]{l@{\hskip 0.4in}c@{\hskip 0.15in}c@{\hskip 0.15in}c@{\hskip 0.4in}c@{\hskip 0.15in}c@{\hskip 0.15in}c}
    Relative error & \multicolumn{3}{c}{$f_{0} \rightarrow f_{1}$} & \multicolumn{3}{c}{$f_{0} \rightarrow f_{2}$}\\
    $[(\Upsilon^{\text{QA}} - \Upsilon^{\text{NR}})\slash\Upsilon^{\text{NR}}]$               & $++$ & $+-$ & $+0$ & $++$ & $+-$ & $+0$ \\
    \midrule
    Time & $\SI{22}{\percent}-\SI{26}{\percent}$ & $\SI{15}{\percent}-\SI{24}{\percent}$ & $\SI{20}{\percent}-\SI{24}{\percent}$ & $\SI{26}{\percent}-\SI{29}{\percent}$ & $\SI{18}{\percent}-\SI{27}{\percent}$ & $\SI{24}{\percent}-\SI{27}{\percent}$ \\
    Phase & $\SI{20}{\percent}-\SI{32}{\percent}$ & $\SI{18}{\percent}-\SI{30}{\percent}$ & $\SI{17}{\percent}-\SI{30}{\percent}$ & $\SI{20}{\percent}-\SI{32}{\percent}$ & $\SI{18}{\percent}-\SI{30}{\percent}$ & $\SI{17}{\percent}-\SI{30}{\percent}$ \\
    GW cycles & $\SI{24}{\percent}-\SI{26}{\percent}$ & $\SI{15}{\percent}-\SI{25}{\percent}$ & $\SI{21}{\percent}-\SI{25}{\percent}$ & $\SI{24}{\percent}-\SI{26}{\percent}$ & $\SI{15}{\percent}-\SI{25}{\percent}$ & $\SI{21}{\percent}-\SI{25}{\percent}$  \\
    GW energy & $\SI{47}{\percent}-\SI{48}{\percent}$ & $\SI{34}{\percent}-\SI{47}{\percent}$ & $\SI{42}{\percent}-\SI{47}{\percent}$ & $\SI{56}{\percent}-\SI{59}{\percent}$ & $\SI{42}{\percent}-\SI{56}{\percent}$ & $\SI{52}{\percent}-\SI{56}{\percent}$    \\
    EM energy & $\SI{42}{\percent}-\SI{44}{\percent}$ & $\SI{59}{\percent}-\SI{76}{\percent}$ & $\SI{70}{\percent}-\SI{76}{\percent}$ & $\SI{47}{\percent}-\SI{51}{\percent}$ & $\SI{76}{\percent}-\SI{99}{\percent}$ & $\SI{91}{\percent}-\SI{98}{\percent}$  \\
    GW angular momentum & $\SI{45}{\percent}-\SI{48}{\percent}$ & $\SI{35}{\percent}-\SI{45}{\percent}$ & $\SI{41}{\percent}-\SI{46}{\percent}$ & $\SI{54}{\percent}-\SI{57}{\percent}$ & $\SI{42}{\percent}-\SI{54}{\percent}$ &  $\SI{51}{\percent}-\SI{55}{\percent}$ \\
    EM angular momentum & $-$ & $\SI{57}{\percent}-\SI{70}{\percent}$ & $\SI{53}{\percent}-\SI{58}{\percent}$ & $-$ & $\SI{71}{\percent}-\SI{87}{\percent}$ & $\SI{65}{\percent}-\SI{69}{\percent}$   \\
    GW150914-like SNR & $\SI{13}{\percent}-\SI{23}{\percent}$ & $\SI{23}{\percent}-\SI{26}{\percent}$ & $\SI{21}{\percent}-\SI{23}{\percent}$ & $\SI{20}{\percent}-\SI{28}{\percent}$ & $\SI{65}{\percent}-\SI{69}{\percent}$  & $\SI{19}{\percent}-\SI{21}{\percent}$  \\
  \end{tabular}
\end{table*}
\end{center}

\subsection{Up to merger}
\label{sec:up-merger}

\begin{figure}[htbp]
  \centering
  \includegraphics{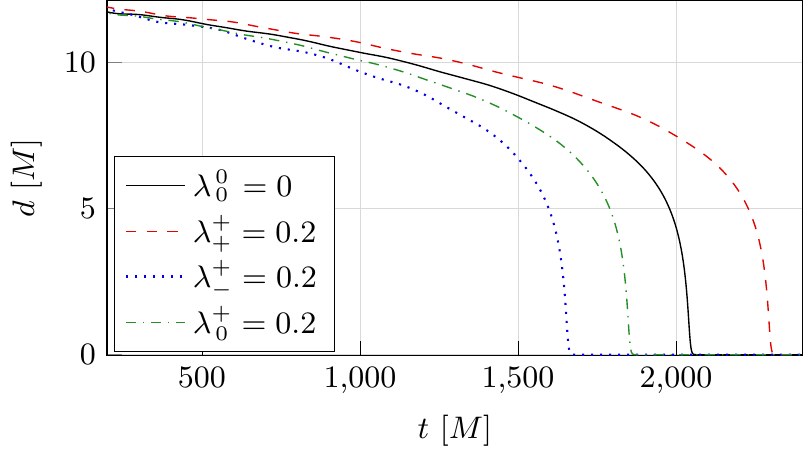}
  \caption{Coordinate distance of the two black holes as a function of the
    coordinate time for three representative cases with charge-to-mass ratio of
    $\lambda = 0.2$, along with the case with no electromagnetic fields. Note that this is not a gauge-independent plot.}
  \label{fig:distances}
\end{figure}

Our simulations capture all non-linear effects that take place during the late
inspiral and merger. So, we can study interesting quantities that are not
accessible with approximate methods. In Fig.~\ref{fig:distances}, we show the
coordinate distance of the two black hole centroids as a function of the
coordinate time for four representative NR simulations. This figure complements
the top panel of Fig.~\ref{fig:eccentricity}. It can be seen that our Newtonian
expectations are met: the system that merges faster is the one with opposite
charge, due to the additional electrostatic attraction and loss of energy due to
dipole electromagnetic emission. Next, we have the one with only one charged
black hole, due to additional loss of energy in the electromagnetic emission.
Finally, the system with black holes with the same charge is the last to merge,
as it has to fight against additional electrostatic repulsion.

An important first finding is that for a fixed binary mass corresponding to
GW150914, the SNR and the number of in-band gravitational-wave cycles depend
very weakly on the charge: $\lesssim \SI{3}{\percent}$. This is in spite of the
different evolutions depicted in Fig.~\ref{fig:distances}. This result will
likely change if we consider low-mass binaries, which have a much longer
inspiral that can be significantly affected by the presence of charge earlier in
the inspiral.

\subsubsection{Properties of the emission}
\label{sec:properties-emission}

In this subsection, we explore some properties of the electromagnetic and
gravitational emission. We find that in our simulations several quantities of
interest scale with $\lambda^{2}$. This scaling is expected to be exact in the
limit where the charge has a negligible contribution to the dynamics, and it
will likely stop to be valid for larger values of $\lambda$, where the
self-gravity of the electromagnetic fields becomes more important. Even if we do
not expect these relations to hold, they are still useful because the
charge-to-mass ratios explored here are up to \num{0.3}, which is a considerable
amount of charge (corresponding to \SI{30}{\percent} of the maximum value
allowed for a general-relativistic black hole).

\begin{figure}[htbp]
  \centering
  \includegraphics{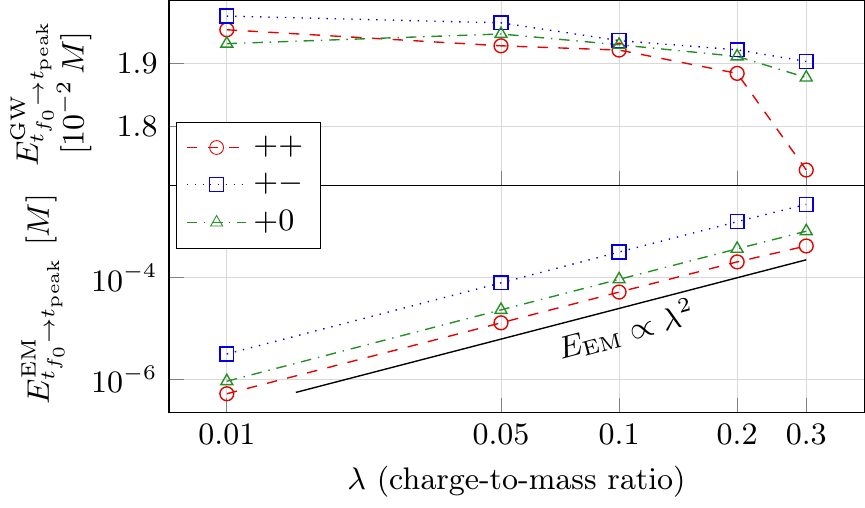}
  \caption{Top: total energy radiated away in gravitational waves (between
    frequencies $f_0$ and $f_{\rm peak}$) as a function of $\lambda$ for the
    different cases we simulated. Bottom: total energy radiated away in
    electromagnetic waves $E_{\text{EM}}$ as a function of $\lambda$. We find
    that the energy scales as the charge squared for for the cases considered in
    this work. The simulation that radiates the most is the one with opposite
    charges because its electric dipole is the largest. In the same-charge case,
    the emission is dominated by the electric quadrupole and it is still
    significant. The angular momentum lost by waves behaves similarly.}
  \label{fig:energy}
\end{figure}

First, we notice that the energy and angular momentum lost in gravitational-wave
emission depend only weakly on the value of $\lambda$ for the cases considered
in our set.  This is shown in the top panel of Fig.~\ref{fig:energy}, where we
plot the total energy lost by gravitational waves for the various simulations we
performed.

Next, it is interesting to study which modes dominate the emission. We focus on
the $\lambda=0.3$ simulations, and begin with the electromagnetic sector. Based
on Newtonian arguments, we expect that in black hole binaries with opposite
charge, most of the emission will be dipolar. By contrast, for simulations with
like charge we expect no dipole, overall reduced emission, mostly in the
quadrupole. Finally, the case with only one charged black hole (which has a
non-zero dipole) should lie in between. The bottom panel of
Fig.~\ref{fig:energy} shows the total energy carried away by electromagnetic
waves starting at the reference frequency $f_0$ until peak gravitational wave
amplitude, and demonstrates that the aforementioned intuition is
correct. Moreover, the figure shows that the energy emitted scales with the
charge-to-mass ratio squared. When computing the mode-by-mode contributions to
the electromagnetic emission, we find that for the $\lambda^{+}_{-}$ case,
$\ell=1$ dominates over all other modes and constitutes \SI{98}{\percent} of the
energy lost. For $\lambda^{+}_{+}$ the $\ell=2$ mode is the main channel through
which energy is carried away, and \SI{98}{\percent} is lost in this way. In the
$\lambda^{+}_{0}$ case both dipole and quadrupole are important, the first
contributing \SI{78}{\percent}, and the second \SI{20}{\percent}. Higher order
modes constitute less than \SI{2}{\percent} contribution. In all of our
simulations the $\ell=2$ mode accounts for \SI{98}{\percent} of the energy lost by
gravitational waves. If we considered angular momentum instead of energy, we
arrive at similar conclusions.

\begin{figure}[htbp]
  \centering
  \includegraphics{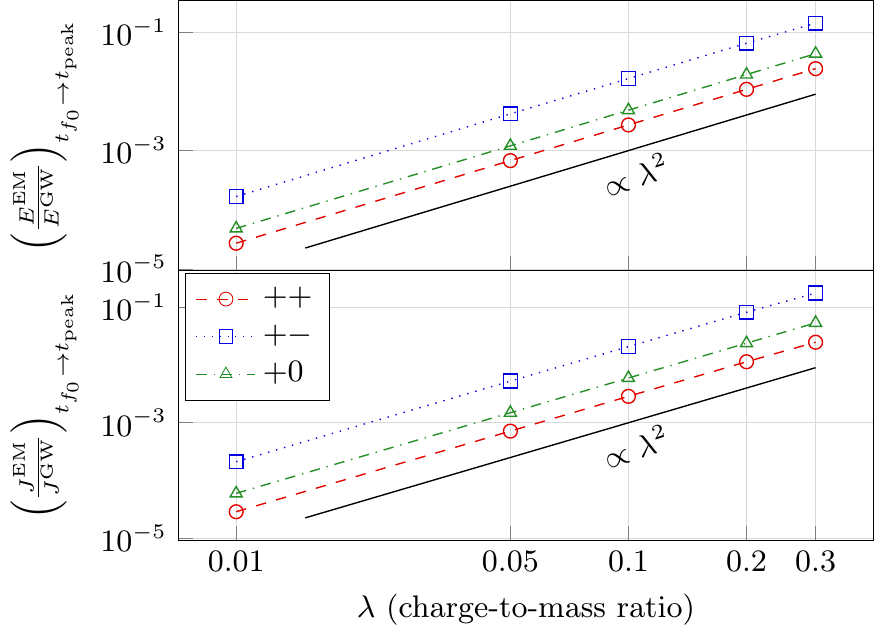}
  \caption{Ratio between electromagnetic and gravitational energy and angular
    momentum lost via emission of radiation from $t_{f_{0}}$ to merger. The
    configuration with the strongest electromagnetic emission is the one with
    opposite charges, as the dipole is the most effective as emitting.}
  \label{fig:ratios}
\end{figure}

The bottom panel in Fig.~\ref{fig:energy} demonstrates that the emitted
electromagnetic energy scales as $\lambda^{2}$. That is not the only quantity
that does so. We find that several quantities scale the same way, including
$E_{\text{EM}}$, $J_{\text{EM}}$, $E_{\text{EM}}\slash E_{\text{GW}}$, and
$J_{\text{EM}}\slash J_{\text{GW}}$. We observe an example of this in
Fig.~\ref{fig:ratios} which shows $E_{\text{EM}}\slash E_{\text{GW}}$, and
$J_{\text{EM}}\slash J_{\text{GW}}$ vs $\lambda$. As remarked at the beginning
of this subsection, while we find the scaling with $\lambda^{2}$ for several
quantities, it is possible that some of this relationships will fail for larger
values of the charge.

In~\cite{Cardoso2016b} it was estimated that for equal-mass mergers and black
holes endowed with opposite charges,
\begin{equation}
  \label{eq:cardoso-eem-egw}
  E^{+-}_{\text{EM}}\slash E^{+-}_{\text{GW}}  \approx 2 \lambda^{2}\,.
\end{equation}
Using our results, which strictly speaking do not apply to the equal mass case
but should be close enough, we find
\begin{equation}
  \label{eq:our-eem-egw}
  E^{+-}_{\text{EM}}\slash E^{+-}_{\text{GW}}  \approx 1.6 \lambda^{2}\,.
\end{equation}
Thus, in spite of the simple derivation in~\cite{Cardoso2016b}, the expression
provided finds the correct order of magnitude in the numerical coefficient.

If we consider charge in its traditional meaning of electric charge, we can talk
about electromagnetic luminosity emitted. For the simulations with $\lambda =
0.3$, the peak electromagnetic luminosity varies from \num{2.7e-5} to
\num{8.6e-5}, corresponding to $\approx \SI{e57}{\erg\per\s}$ (regardless of the
value of \si{\admmass}). For $\si{\admmass} = \SI{65}{\sunmass}$, the maximum
electromagnetic energy emitted is $\approx \SI{e53}{\erg}$, which is much
smaller than the energy emitted through gravitational waves. This possibly
explains the near perfect scaling with $\lambda^2$ of several quantities we
reported. How this electromagnetic energy is converted to potentially observable
photons has to be modeled and it is highly dependent on the environment. In
standard astrophysical conditions, the diluted plasma would not allow these
waves to propagate~\cite{Cardoso2020}.

\subsubsection{Charge detectability}
\label{sec:detect-charge-with}

In this subsection, we use our simulation results to estimate black hole charge
detectability by future gravitational wave detectors. To do this, we
follow~\cite{Bozzola2020} and compute the mismatch between two signals $h_1$ and
$h_2$. This quantity determines the minimum SNR needed to distinguish $h_{1}$
and $h_{2}$, which we denote as SNR$(h_{1}, h_{2})$, and is given
by~\cite{Flanagan1998b,Lindblom2008,McWilliams2010,Baird2013,Abbott2017h}
\begin{equation}
  \label{eq:snr-mism}
  \text{SNR} (h_{1}, h_{2}) = \frac{1}{\sqrt{2}} \frac{1}{\sqrt{\text{mismatch}(h_{1}, h_{2})}}\,.
\end{equation}
Note that the SNR threshold value of Eq.~\eqref{eq:snr-mism} corresponds to a
\SI{68}{\percent} confidence level for distinguishability. We compute the mismatch
as~\cite{Damour1998, Abbott2017h, Bozzola2020}
\begin{equation}
  \label{eq:mismatch}
  \text{mismatch}(h_1, h_2) = 1 - \max \mathcal{O}(h_1, h_2)\,,
\end{equation}
with the maximum evaluated with respect to time-shifts, polarization angles, and
mass shifts. We maximize over mass shifts to include the possible degeneracy
between mass and charge (see~\cite{Bozzola2020} for a detailed discussion). This
operation is allowed because our simulations scale with the total mass
$\si{\admmass}$. In Eq.~\eqref{eq:mismatch}, $\mathcal{O}(h_1, h_2)$ is the
overlap between $h_1$ and $h_2$
\begin{equation}
  \label{eq:overlap}
  \mathcal{O}(h_1, h_2) = \frac{(h_1, h_2)}{\sqrt{(h_1, h_1) (h_2, h_2)}}\,,
\end{equation}
with $(h_1, h_2)$ being the noise-weighted inner product between the two signals
in the frequency domain $\tilde{h}_1(f)$ and $\tilde{h}_2(f)$, which is given
by~\cite{Harry2011}
\begin{equation}
  \label{eq:inner-produced}
  (h_1, h_2) = 4 \mathrm{Re } \int_{f_{\text{min}}}^{f_{\text{max}}}
    \frac{\tilde{h}_1(f) \tilde{h}_2^{\star}(f)}{S_n(f)} \d f \,,
\end{equation}
where $S_n(f)$ is the power spectral density of the detector noise, and the
asterisk indicates complex conjugation. In the subsequent analysis we consider
the following detectors at design sensitivity: Advanced LIGO,
A+~\cite{Miller2015}, Voyager~\cite{Voyager}, Einstein
Telescope~\cite{Punturo2010}, and Cosmic Explorer~\cite{CosmicExplorer} and we
set $f_{\text{min}} = \SI{25}{\Hz}$ and
$f_{\text{max}} = \SI{1024}{\Hz}$.\footnote{When available, we use the
  sensitivity curves implemented in LALSimulation~\cite{lalsuite} via
  PyCBC~\cite{PyCBC}. The functions used are part of the \texttt{pycbc.psd}
  module and are: \texttt{aLIGODesignSensitivityP1200087},
  \texttt{aLIGOAPlusDesignSensitivityT1800042},
  \texttt{EinsteinTelescopeP1600143}, and \texttt{CosmicExplorerP1600143}. For
  Voyager we obtained the sensitivity curve from LIGO Document
  T1500293-v11. \label{fn:psd}}. Note that here we are not including the effects
of spins, mass-ratios, and eccentricity, parameters that introduce degeneracies
that make the detection of charge more difficult. Nonetheless, the key result of
this subsection will hold at a more qualitative level even when the additional
parameters are included: for a GW510914-like event, future instruments are
expected to detect signals with a SNR significantly larger than the one needed
to detect charge at the level of $\lambda \approx \num{0.1}$. For precision
studies, more accurate simulations that include also the other parameters are
needed.

Before we discuss the results, we highlight possible pitfalls in numerically
evaluating Eq.~\eqref{eq:inner-produced}. There are multiple steps to go from
what the simulations output to Eq.~\eqref{eq:inner-produced}. In particular,
Fourier transforms have to be computed and the time series have to be windowed
and zero-padded to avoid aliasing and spectral leakage. Since the simulations we
are considering produce similar waveforms (up to time, phase, and mass shifts),
small differences in how the two waves are pre-processed can contribute
significantly to the value of the mismatch. First, it is important to trim the
end of the two waveforms where the strain is practically zero and ensure that
they have the same duration after the peak. If this is not done, and the signals
have different durations after merger, applying a window has a different effect
and introduces a systematic uncertainty. Second, in some cases the fixed
frequency integration may leave small residual drifts at the very end of the
waveform. These depend on the simulation, so one must check that all the waves
are well-behaved. In case they are not, one can adjust this by cropping the
signal or by changing the parameters of the window function or of the
integration method. Improperly considering one of these effects may lead to
systematic errors.

\begin{figure}[htbp]
  \centering
  \includegraphics{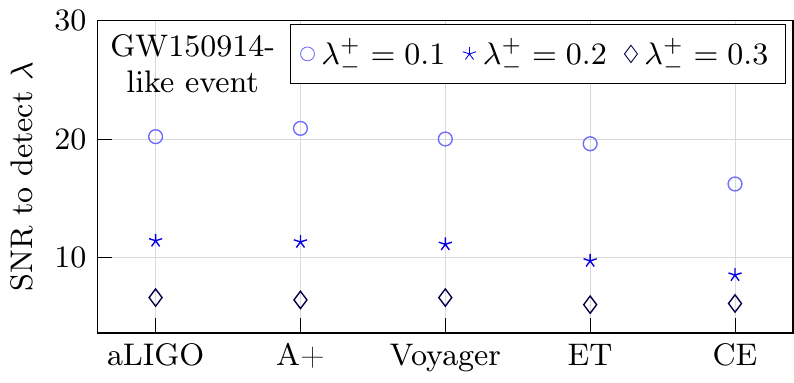}
  \caption{Signal-to-noise ratio required to distinguish waveforms from mergers
    of charged black holes from ones without charge for different detectors
    assuming their design sensitivity. The expected SNR for a GW150914-like
    event future detectors is significantly larger than the one needed to
    constrain $\lambda^{+}_{-} < 0.1$. See Footnote~\ref{fn:psd} for details on
    the detector sensitivity curves used.}
  \label{fig:constraints}
\end{figure}

We now discuss the results of our study by focusing on the $\lambda^{+}_{-}$ case,
which is relevant to constraining the dipole emission. In
Fig.~\ref{fig:constraints}, we report the value of the SNR needed to distinguish
uncharged binary waveforms from binary black holes with charge-to-mass ratio
$\lambda^{+}_{-} = 0.1$ (circles), $0.2$ (stars), or $0.3$ (diamonds) for a
GW150914-like event. For smaller values of $\lambda^{+}_{-}$, the mismatch computed
from our simulations is limited by their numerical error, so higher resolution
evolutions would be needed.

First, we note that there is not much variation of the SNR for
distinguishability across the different detectors, regardless of the significant
variation in sensitivity. The reason for this result is that for the computation
of the mismatch the overall noise curve does not matter: it is how the noise is
distributed in different frequencies that matters the most. The difference in
sensitivity is reflected in how easy or not it is to achieve such SNRs. Second,
the values needed to detect $\lambda^{+}_{-} \geq 0.1$ are already achievable today
(GW150914 had a network-SNR of approximately 25~\cite{Abbott2016d}, but its
noise curve was not the one at deign sensitivity). Given their improved
sensitivity, future detectors will immediately be able to detect this amount of
charge. To estimate what limits on charge future detectors will place, one not
only needs better simulations, but one needs to include the effects of spin and
eccentricity.

\begin{figure}[htbp]
  \centering
  \includegraphics{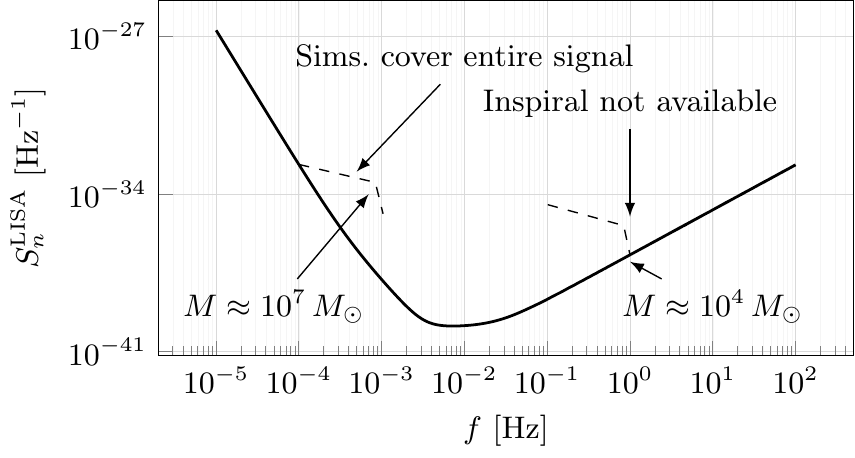}
  \caption{Schematic representation of how we can use our simulations for LISA
    sources. We are free to vary the mass, but in many cases our simulations do
    not cover the entirety of the signal, and part of the inspiral is missing.
    Since the inspiral is the most constraining part of the signal, when we
    compute the minimum SNR for distinguishability of charged binary waveforms
    from uncharged binary ones, we are providing an upper bound.}
  \label{fig:lisa}
\end{figure}

Next, we discuss what charge-to-mass ratio LISA would be able to detect for
million solar mass binaries. As already mentioned, we are free to rescale the
mass of our binaries to place them in the LISA band, which we assume ranges from
\SI{0.1}{\milli\Hz} to \SI{1}{\Hz}. Despite the scale freedom, there is one
obstacle to doing a complete analysis using our simulation data as
Fig.~\ref{fig:lisa} demonstrates. In the plot, we show the LISA sensitivity
curve~\cite{Robson2019} and we schematically show with dashed lines the
gravitational wave spectrum from our simulations when the mass is rescaled using
two different values. In the case of a $10^7M_\odot$ binary, our simulation
signal is entirely in the LISA band. On the other hand, at least part of the
inspiral is missing for $\sim 10^{4-6}M_\odot$ binaries.\footnote{One could
  extended our simulation data with some approximate waveforms, like the one
  from the Newtonian model, and recompute the quantities with the entire
  signal. However, this goes beyond the scope of the current work.} Despite this
obstacle, we know that for the case with opposite charge, the mismatch will
increase if we included the inspiral, due to the presence of dipole
emission. Hence, if we use our data to find what is the minimum SNR needed to
distinguish waveforms of charged binaries from those generated by uncharged
binaries, we will find an upper bound on that (if we included the entire in-band
inspiral waveform the minimum SNR for distinguishability would only decrease).

\begin{figure}[htbp]
  \centering
  \includegraphics{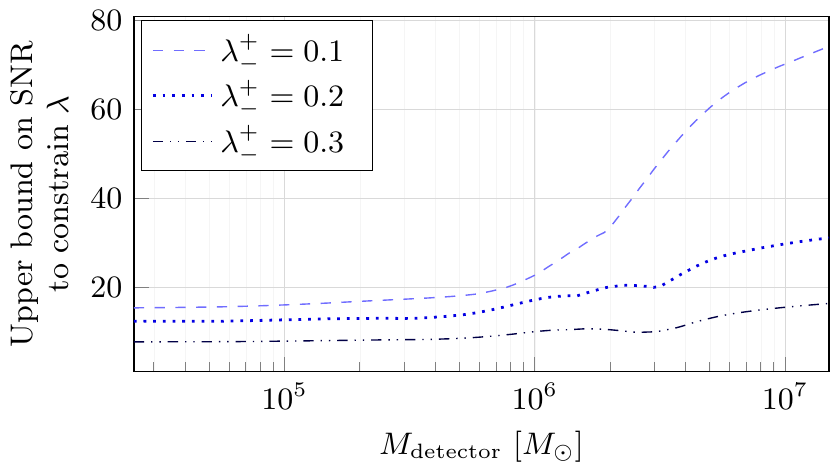}
  \caption{Upper limit on the minimum SNR needed to distinguish waveform from
    charged binaries with different $\lambda^{+}_{-}$ from uncharted systems.
    This is an upper limit because most of the masses do not include the
    entirety of the inspiral. Given that the inspiral is most constraining,
    including it would decrease the SNR needed.}
  \label{fig:lisa_constrain}
\end{figure}

Figure~\ref{fig:lisa_constrain} shows the upper bound on the minimum SNR needed
to distinguish waveforms generated by charged binaries from those generated by
uncharged binaries for different charge configurations as a function of the
detector-frame mass $M_{\text{detector}}$. LISA is expected to detect binaries
with SNR much higher than the one in Fig.~\ref{fig:lisa_constrain}, so it will
be able to detect or place constraints on small values of $\lambda^{+}_{-}$ for
multiple systems. These results are not surprising. In~\cite{Cardoso2016b,
  Cardoso2020Erratum} it argued that LISA will constrain the dipole moment at
the level of the \num{e-4}, considering only the inspiral.

We can understand the shape in Fig.~\ref{fig:lisa_constrain} by considering that
the inspiral has the most important contribution to the mismatch, so, it is the
lower frequencies that matter the most. Recall that Fig.~\ref{fig:lisa} shows
schematically the power spectrum of the gravitational waves. Roughly speaking,
there is more power in the lower frequencies than the higher ones because more
time is spent there. Increasing the binary mass from the minimum value
considered here amounts to sliding the spectrum in Fig.~\ref{fig:lisa} from the
right to the left. When we are considering masses for which the signal lies to
the right-hand-side of the plot, we find that there is more noise in the higher
frequencies of the signal than the lower ones. This is the optimal condition to
separate an uncharged waveform from a charged ones, because it is the lower
frequencies that contain the most information. Hence, the signal to noise for
the detection is the lowest. Now, when we consider masses for which the signal
lies on the left-hand-side of Fig.~\ref{fig:lisa}, initially not much changes in
the signal-to-noise ratio. This is because the frequencies with the most
information are the ones with the lowest noise. The scenario changes when we
approach the minimum of the sensitivity curve, at that point, the way noise is
distributed across frequency changes, and we see in
Fig.~\ref{fig:lisa_constrain} an increase in the signal-to-noise ratio, which
reflects the fact that we are removing sensitivity in the lower frequencies to
put it in the higher ones (that have less information). The trend continues when
we increase further the mass scale and we climb up the sensitivity curve on the
left side. Here, we are giving more weight to high frequencies, which cannot
distinguish the two waves well, so we need more signal-to-noise ratio for
distinguishability of two signals.

\begin{figure}[htbp]
  \centering
  \includegraphics{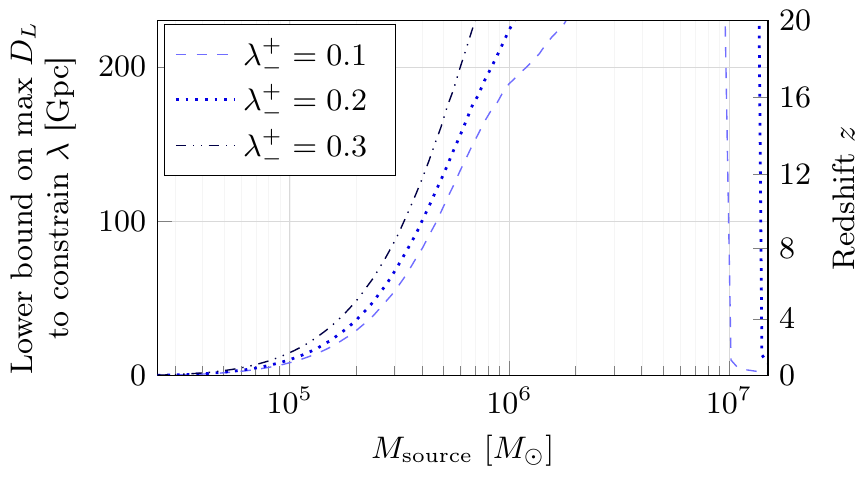}
  \caption{Lower limit on the maximum luminosity distance $D_{L}$ in gigaparsec
    (or max redshift $z$) at which LISA can detect a charged binary with
    $\lambda$ of 0.1 as function of the source-frame mass $M_{\text{source}}$.}
  \label{fig:lisa_distance}
\end{figure}

A natural question that arises next is the following: how far can LISA detect
the minimum SNR required to distinguish charge? Computing the SNR with
Eq.~\eqref{eq:snr}, we can find what is the maximum distance at which the SNR is
larger than the threshold. This distance is plotted in
Fig.~\ref{fig:lisa_distance}. For some mass ranges, LISA will essentially
distinguish a charged binary with $\lambda^+_-$ up to 0.1 everywhere it can detect
black hole binaries. It is important to note that that plot provides only a
lower limit on this maximum distance, especially for lower masses, where we do
not include the long inspiral part of the waveform.

We can ask the same question for the case $\lambda^{+}_{+} = 0.3$ at the mass-scale
for which we have the entire signal in LISA band ($\approx \SI{e7}{\sunmass}$). We
find that LISA will be able to detect charge up to redshift of about 1, which
would translate to a constraint on the $\alpha$ parameter of Moffat's
Scalar-Vector-Tensor gravity~\cite{Moffat2006} of $\alpha \lesssim 0.1$ (see
also,~\cite{Bozzola2020}).

\subsection{Properties of the remnant black hole and ringdown}
\label{sec:prop-final-black}

Here we study the properties of the remnant black hole forming following the
merger of charged binaries. We discuss the accuracy of the method presented in
Sec.~\ref{sec:spin} to estimate the remnant black hole spin
(Sec.~\ref{sec:final-spin}), and we explore the quasi-normal-modes from the
ringdown phase (Sec.~\ref{sec:quasi-normal-modes-1}).

\subsubsection{Remnant black hole properties}
\label{sec:final-spin}

In this subsection, we discuss the physical properties of the charged binary
black hole merger remnants in our numerical-relativity simulations as computed
via \texttt{QuasiLocalMeasuresEM}. Then, we compare the remnant black hole spin
found in the simulations with the expectations of the BKL approximation
described in Sec.~\ref{sec:spin}.

Before presenting our results, let us consider what one might expect
qualitatively. Following the discussion in Sec.~\ref{sec:contr-eccentr}, we know
that, in Newtonian physics, the orbital angular momentum scales as
$\sqrt{G}$. Next, consider conservation of angular momentum, and assume (as in
the BKL method) that the spin of the final black hole is determined by the
angular momentum at the innermost-stable circular orbit of the remnant black
hole. We would expect the final spin $\chi^{\text{final}}$ with respect to the
uncharged case $\chi^{\text{final}}_{\text{00}}$ to increase as $\sqrt{1 -
  \lambda_{1}\lambda_{2}}$ for the case with opposite charge (where the ISCO of
the remnant black hole is the same as in the uncharged case\footnote{Although
  the ISCO of the remnant black hole in the charged case with opposite charges
  is practically unaffected by charge (for the range of $\lambda$ in this work),
  since the total charge is approximately zero, the effective innermost stable
  orbit of the binary (the orbital separation at plunge) must be affected to
  some extent, especially for larger values of $\lambda$ even when the component
  black holes have exactly opposite charges.}), and to decrease faster than
$\sqrt{1 - \lambda_{1}\lambda_{2}}$ for the other two cases, where the ISCO is
reduced compared to the uncharged case, resulting in additional emission of
angular momentum.

\begin{figure}[htbp]
  \centering
  \includegraphics{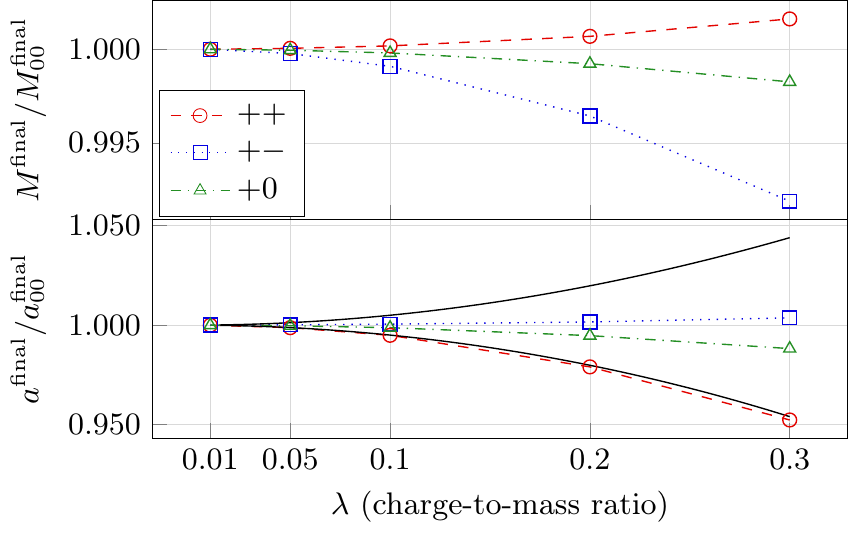}
  \caption{Mass (top panel) and spin (bottom panel) of the remnant black hole
    computed with the isolated horizon formalism~\cite{Ashtekar2000,
      Ashtekar2001, Ashtekar2004} as implemented in
    \texttt{QuasiLocalMeasuresEM}~\cite{Bozzola2019}, compared to the one in a
    merger with no charge ($M^{\text{final}}_{00} \approx \SI{0.96}{\admmass}$,
    $a^{\text{final}}_{00} \approx \SI{0.66}{\admmass}$). The solid black lines in the
    bottom panel indicate the Newtonian scaling $\sqrt{1\pm\lambda_1\lambda_2}$. }
  \label{fig:finalbh}
\end{figure}

In Fig.~\ref{fig:finalbh} we show how these properties change with respect to
their values in the uncharged case. First, we find that the final mass of the
remnant is almost independent of the charge configuration, with sub-percent
variations. In particular, the $\lambda^{+}_{-}$ remnant looks almost identical
to the one from the uncharged simulation: spin, charge, and mass are the same to
within \SI{1}{\percent}. Moreover, in that case, the angular momentum is
essentially independent of $\lambda$, against the expectations from the
Newtonian argument. In general, we find that the final properties of the black
hole depend weakly on the charge configuration, with the largest variation being
about \SI{4}{\percent} in the spin of the $\lambda^{+}_{+} =0.3$
case. Interestingly, the cases with like charge follow the Newtonian scaling
$\sqrt{1-\lambda_1\lambda_2}$ as shown in Fig.~\ref{fig:finalbh} with the black
solid lines.

\begin{figure}[htbp]
  \centering
  \includegraphics{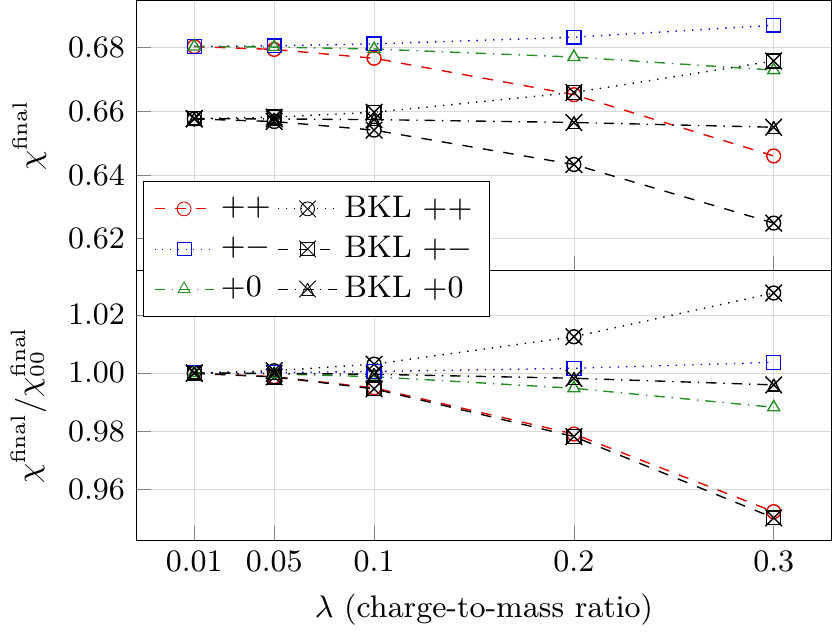}
  \caption{Dimensionless spin of the remnant black hole in the numerical
    relativity simulations (colored lines) and estimated with the method of
    Sec.~\ref{sec:final-black-hole} (black lines with crossed markers). In
    the top panel, we compare the actual numerical values, in the bottom panel
    we rescale the values by the value measured in the simulation without charge
    or the estimated value in absence of charge (simulation data is normalized
    with the value obtained from the uncharged simulation, and BKL estimates are
    normalized with the value obtained assuming no charge).}
  \label{fig:final_spin}
\end{figure}

In Sec.~\ref{sec:spin}, we described a simple way to estimate the spin of the
remnant using conservation arguments. We now compare the method with our
simulations. The top panel of Fig.~\ref{fig:final_spin} shows that the method
estimates the value of the final dimensionless spin with an error of \SI{3}{\percent}
in the uncharged case. This method does not include energy loss. Including it
would not improve accuracy, as the dimensionless spin would be overestimated as
opposed to underestimated. In the bottom panel of Fig.~\ref{fig:final_spin} we
normalize the final spin to the value of the uncharged case with simulations
normalized with the values obtained from the simulations, and the BKL values are
normalized with the value obtained applying the method in absence of charge.
This removes the normalization as a possible parameter and allows us to test if
the method captures the correct trend (and only gets the normalization wrong).
In practice, this corresponds to testing whether removing trends arising from
the normalization makes the black lines in the bottom panel of
Fig.~\ref{fig:final_spin} overlap with the colored ones. We observe that the
method indeed works well for the cases with like charge or with only one charged
black hole, but overestimates the final spin in the other cases. The approach
does not capture the fact that the spin of the final black hole in the
$\lambda^{+}_{-}$ numerical relativity simulations is essentially independent of
charge. However, the method is still accurate to within a few per cent.

\subsubsection{Quasi-normal-modes}
\label{sec:quasi-normal-modes-1}

Here we discuss the ring-down phase and address how challenging it is to tell if
a binary black hole is charged from the ringdown phase alone.

\begin{figure}[htbp]
  \centering
  \includegraphics{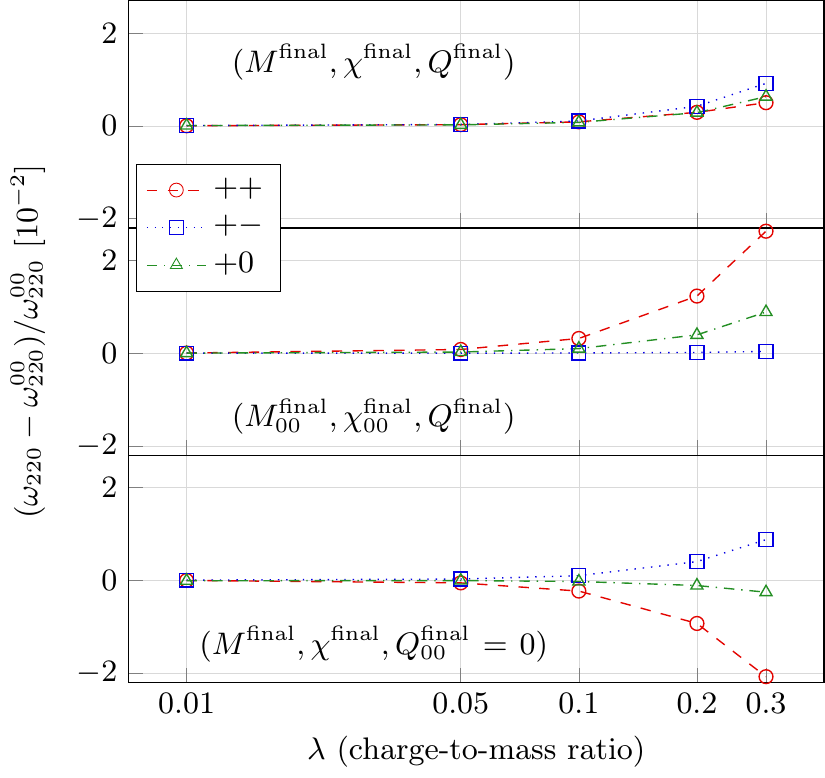}
  \caption{Relative difference between the analytical 220 quasi-normal-mode
    frequency for the merger remnants in our evolutions and the simulation with
    uncharged black holes. In the middle panel, we remove the effects of mass
    and spin and fix the values to the ones they have in the uncharged case. In
    the bottom panel, we remove the contribution of the charge described by
    Eqs.~\eqref{eq:qnm-kn}: the frequency becomes more different. The same
    happens with the decay time, but the variation with respect to the uncharged
    case is even smaller. }
  \label{fig:qnm_total}
\end{figure}

Since the frequency and decay time of the quasi-normal-modes depend only on
mass, spin, and charge, we can use the values reported in Fig.~\ref{fig:finalbh}
to compute $\omega_{220}$ and $\tau_{220}$ with Eqs.~\eqref{eq:qnm-kn} and the
\texttt{qnm} code~\cite{Stein:2019mop}. The top panel of
Fig.~\ref{fig:qnm_total} shows the relative difference of the computed value
$(\omega_{220}-\omega_{220}^{00})/\omega_{220}^{00}$, where $\omega_{220}^{00}$
is the ring-down frequency computed with {\tt qnm} using the mass and spin of
the remnant black hole from the simulation without charge, while $\omega_{220}$
is the ring-down frequency using Eqs.~\eqref{eq:qnm-kn}. The maximum difference
we find is \SI{0.9}{\percent} in the $\lambda^{+}_{-} = 0.3$ case. Therefore, to
be able to distinguish the ringdown from a merger of charged black holes from
one with no charge, the parameters have to be estimated better than the percent
level. For a given value of $\lambda$, even higher accuracy is needed to be able
to distinguish the three charged scenarios.

It is interesting to understand why the simulation with opposite charge, in
which the remnant has the least amount of total charge for a fixed $\lambda$, is
the one with largest difference with respect to the uncharged case: In
Fig.~\ref{fig:final_spin}, we see that for the case with opposite charges
$\chi^{\text{final}}\slash \chi^{\text{final}}_{00} \approx 1$ regardless of the
value of $\lambda$. Combining this information with the fact that the remnant
black hole has $Q\slash M \ll 1$, we deduce that the only difference between the
uncharged case and the case with $\lambda^{+}_{-} = 0.3$ must be in the mass of
the final black hole. The difference in mass alone is responsible for the
observed difference in the ringdown frequencies.

It is also interesting to understand why the case with same charge is not the
easiest to distinguish, despite that the remnant black hole mass in this case
has $\lambda \approx 0.3$, and is the one with the most different spin
(Fig.~\ref{fig:final_spin}) compared to the uncharged case. Interestingly, in
the $\lambda^{+}_{+}$ case, charge, angular momentum and mass conspire so that the
remnant black hole appears to have a quasi-normal mode frequency that matches
that from an uncharged binary. To analyze this case, it is convenient to
separate the effects of charge, and mass and spin. The fundamental properties of
the quasi-normal-modes, $\omega_{220}$ and $\tau_{220}$, depend on the triplet
$(M^{\text{final}},\chi^{\text{final}},Q^{\text{final}})$. We can hold some of
these parameters fixed to the case with no charges to understand what is the
dominant contribution to the deviation of $\omega_{220}$ and $\tau_{220}$ from their
corresponding values in the absence of charge. The results are reported in the
middle and bottom panels of Fig.~\ref{fig:qnm_total}, where we report $\omega_{220}$.
The same results hold for $\tau_{220}$.

In the middle panel we vary only the charge, and fix the mass and spin to the
value they obtain in the simulation with no charge $M^{\text{final}}_{00}$ and
$\chi^{\text{final}}_{00}$. Here we find that binaries with opposite charge have
remnant black holes that behave almost exactly like the uncharged one (there is
no difference in the quasi-normal-modes). This tells us that for this specific
configuration the deviation in $\omega_{220}$ is not because of charge, but that the
mass and spin are the parameters that primarily control the quasi-normal-modes
properties. This is not unexpected: oppositely charged binaries are those with
the smallest remnant black hole charge, so it is natural that this parameter
will contribute the least if mass and angular momentum are the same. On the
other hand, the case with like charges is the most different from the uncharged
ones, with relative increase in $\omega_{220}$ of \SI{2.6}{\percent}.

The bottom panel of Fig.~\ref{fig:qnm_total} shows what happens when we
completely ignore the contribution of charge to the quasi-normal modes. First,
we find the confirmation that here is where the simulations $\lambda^{+}_{-}$ acquire
the differences of up to $\approx \SI{1}{\percent}$ shown in the top panel of
Fig.~\ref{fig:qnm_total}. Second, in the like charges case, we also find that
charge introduces differences in the mass and angular momentum of the remnant
black hole that drive the change in the quasi-normal mode in the opposite
direction compared to the ones shown in the middle panel. Therefore, when we
consider the complete triplet of mass, spin, and charge (``summing'' middle and
bottom panels), we find that their effects almost cancel each other, so that the
corresponding quasi-normal-modes are close to the uncharged binary case.

Hence, we conclude that for the $\lambda$ explored here, to distinguish the
quasi-normal modes of the final black hole from the ones of an uncharged black
hole, one needs exquisite accuracy. The task becomes even harder when one
includes the other parameters which we kept fixed in our simulation (mass-ratio,
spin, and eccentricity), which will introduce additional degeneracies. The
accuracy needed is also of the same order as that of the fitting functions in
Eq.~\eqref{eq:qnm-kn}, and for the highest values of $\lambda$ adopted here, it
is also of the same order as the errors due to the truncation of the expansion
in $\lambda$, which may affect the result. In our simulations, the ringdown
signals are essentially indistinguishable from one another, and we cannot
identify the quasi-normal-mode parameters at the accuracy needed to tell them
apart. To sum up, values of charge up to $\lambda\sim 0.3$ could be challenging
to detect using the ring-down phase alone.

\section{Conclusions and future directions}
\label{sec:conclusion}

In this paper, we continued our program of exploring the non-linear dynamics of
black holes in Einstein-Maxwell theory. In Sec.~\ref{sec:numerical-relativity},
we described our theoretical and numerical approach, emphasizing new features
and formalism that have not been treated before, including: how to prepare
quasi-circular initial data for charged black hole binaries
(Sec.~\ref{sec:contr-eccentr}), how to perform long-term and stable evolutions
of quasi-circular inspirals of charged black holes
(Sec.~\ref{sec:achieving-long-term}), the electromagnetic contribution to
horizon properties of black holes (Sec.~\ref{sec:contribution-spin-em}), and the
computation of the angular momentum carried away by electromagnetic waves with
the Newman-Penrose formalism (Sec.~\ref{sec:grav-electr-waves}).

We compared the results of our non-linear simulations with approximate
approaches for the inspiral (Sec.~\ref{sec:inspiral}) and the ringdown
(Sec.~\ref{sec:prop-final-black}). For the systems considered, our work shows
that Newtonian models based on the quadrupole approximation find the correct
order of magnitude in a set of gauge-invariant quantities, but have errors
$\mathcal O(20\%)$ or larger, and hence they cannot be used in precision studies
of mergers of charged black holes or accurate parameter estimation. Similarly,
estimates of the spin of the remnant black hole based on conservation of angular
momentum and energy arguments are accurate up to few percent. Hence, a key
result of this work is extending what was found in~\cite{Zilhao2012}: these
arguments can be used to build intuition and make order-of-magnitude estimates
in the case of quasi-circular mergers, too.

Furthermore, we discussed properties of the emission (Sec.~\ref{sec:up-merger})
and estimated the detectability of charge by future gravitational wave
observatories (Sec.~\ref{sec:detect-charge-with}), focusing in particular on
LISA.\@ Finally, we studied the quasi-normal-modes
(Sec.~\ref{sec:quasi-normal-modes-1}), finding that it may be challenging to
extract charge information from the ringdown alone.

There are multiple possible extensions of this work. On the non-linear side,
including spin and increasing the charge-to-mass ratio would allow the
exploration of a region of the parameter space never considered before. On the
side of approximate calculations, the next step in complexity after Newtonian
physics is developing 1PN models. Such waveforms are currently
available~\cite{Julie2018b, Khalil2018}. These models are important in the
effort of generating gravitational-wave templates.

\begin{acknowledgments}
  \small We used \texttt{kuibit}~\cite{kuibit} for part of our analysis. We
  thank M.\ Zilh{\~a}o for help with \texttt{ProcaEvolve}. We are grateful to
  the developers and maintainers of the open-source codes that we used. This
  work was in part supported by NSF Grant PHY-1912619 to the University of
  Arizona. G.\ B.\ is supported by a Texas Advanced Computing Center (TACC)
  Frontera Fellowship. Frontera is supported by NSF grant No.\ OAC-1818253. We
  acknowledge the hospitality of the Kavli Institute for Theoretical Physics
  (KITP), where part of the work was conducted. KITP is partially supported by
  the NSF grant No.\ PHY-1748958. Computational resources were provided by the
  Extreme Science and Engineering Discovery Environment (XSEDE) under grant
  number TG-PHY190020. XSEDE is supported by the NSF grant No.\ ACI-1548562.
  Simulations were performed on \texttt{Comet}, and \texttt{Stampede2}, which is
  funded by the NSF through award ACI-1540931.
\end{acknowledgments}

\appendix

\section{Courant stability of Kreiss-Oliger Dissipation}
\label{sec:dissipation}

Kreiss-Oliger dissipation is described by Eq.~\eqref{eq:diss-1}, which we
rewrite here for convenience
\begin{equation}
  \label{eq:diss-1-2}
  \partial_t U = \cdots + {(-1)}^{(p+3)/2} \frac{\epsilon}{2^{p+1}} \Delta x^{p} \partial^{p+1}_{x}U\,.
\end{equation}
This technique has a free a parameter $\epsilon$ that controls the strength of the
dissipation. This cannot be chosen arbitrarily, as we show in this Appendix.

A finite-difference discretization transforms Eq.~\eqref{eq:diss-1} to
\begin{equation}
  \label{eq:diss-2}
  \partial_t U = \cdots + {(-1)}^{(p+3)/2}  \frac{\epsilon}{2^{p+1}} \Delta x^{p} \frac{D^{p+1} U}{\Delta x^{p+1}}\,,
\end{equation}
where $D^{p+1} U$ is determined by the finite-difference stencil. For an
explicit time integration scheme, this operator alone leads to a Courant
stability condition of the form
\begin{equation}
  \label{eq:diss-cfl}
  \frac{\Delta t}{\Delta x} \frac{\epsilon}{2^{p+1}} \le \Lambda\,,
\end{equation}
where $\Lambda$ is typically a fixed number that is determined by the details of
numerical integration method.\footnote{The case with $p=1$ using standard
  centered second-order accurate finite differences is a textbook example of
  stability analysis for a standard diffusion equation with constant
  coefficients. In this case, and for a three-dimensional Cartesian grid with
  equal grid spacing in the all spatial directions, $\Lambda = 1\slash 6$.} Given that in
our case we add dissipation to hyperbolic partial differential equations,
condition~\eqref{eq:diss-cfl} can be recast to read
\begin{equation}
  \label{eq:diss-cfl2}
  \epsilon \le \frac{\Lambda}{\mu}\,,
\end{equation}
where $\mu$ is the Courant factor $\Delta t \slash \Delta x$. For the Einstein-Maxwell equations
in standard finite-difference implementations, this quantity is typically chosen
such that $\mu \leq 0.5$ for numerical stability. Since we are working with an
adaptive-mesh-refinement computational grid with sub-cycling in time, there are
multiple values of $\Delta t$ and of $\Delta x$, so condition~\eqref{eq:diss-cfl} can be
met in some parts of the grid but not in others. Failure to satisfy the
condition can result in numerical instabilities that spoil the simulation. It
should be noted that the Courant condition~\eqref{eq:diss-cfl2} is necessary but
not sufficient for stability. This means that the scheme can be unstable even
when Eq.~\eqref{eq:diss-cfl2} is satisfied. For example, an ill-posed initial
boundary value problem or other numerical instabilities could cause simulations
to crash.

\section{Parameters of the numerical evolution}
\label{sec:param-test-evol}

In this Appendix, we report the parameters used for our simulations, including
the test described in Sec.~\ref{sec:achieving-long-term}.

Our simulations are on a grid with outer boundary at $\SI{1033}{\admmass}$ and
nine refinement levels with refinement boundaries located at
$2^{i}\SI{}{\admmass}$, where $i$ is the number of level. Our standard
coarsest-level resolution is $\approx \SI{4}{\admmass}$ and the high-resolution runs
are with \SI{25}{\percent} smaller grid spacing ($\approx \SI{3.2}{\admmass}$). Of the nine
refinement levels, five have the same timestep, the other levels had Courant
factor fixed to $\Delta t \slash \Delta x$ of $0.4$. This choice reduces the maximum timestep
on the grid and prevents some numerical instability that would otherwise arise.
We set $\kappa$ (defined in Appendix A in~\cite{Zilhao2015}) to
$\SI{9.9}{\per\admmass}$ and verified that changing this parameter does not
produce significant differences. We set the $\eta$ parameter in the evolution of
the shift vector (as defined in Eq.~(13) in~\cite{Sperhake2007}) to
$\SI{1.5}{\per\admmass}$. These parameters lead to instability unless
$\kappa \lesssim {1.5}\slash {\Delta t_{\text{max}}}$ (same for $\eta$), so we adjusted the time-stepping
in our evolutions to ensure that this condition is met. We use fifth order
prolongation in space and second in time. We tested selected cases with seventh
order spatial prolongation and found no significant differences. All our
derivatives are obtained with sixth-order finite difference.

The test described in Sec.~\ref{sec:achieving-long-term} was performed with the
high-resolution grid and otherwise the same parameters described above. The
tests showed that the instabilities arise unless we choose the continuous
prescription for the dissipation. In the unstable cases, we found that
simulations with higher resolution become unstable earlier compared to the ones
at lower resolution.

\begin{figure}[htbp]
  \centering
  \includegraphics{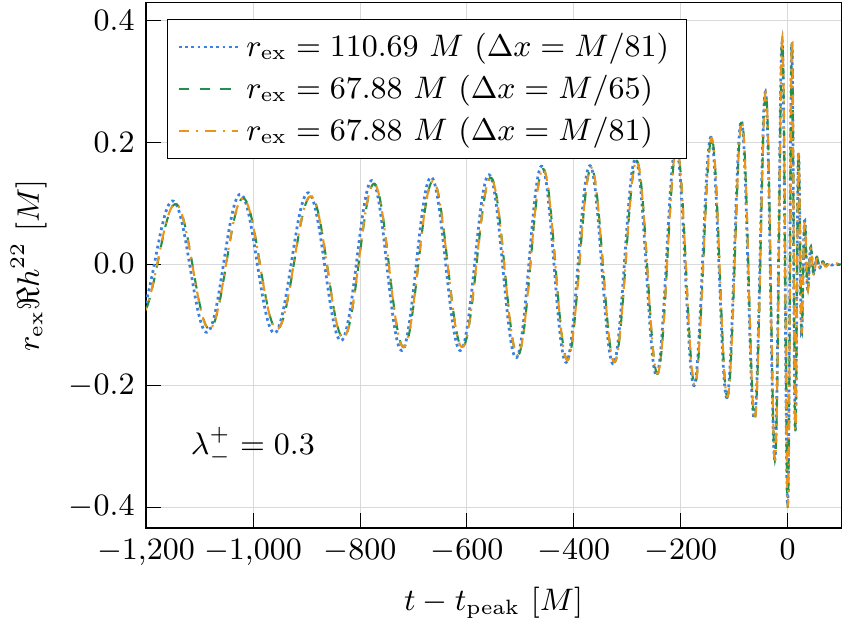}
  \caption{Real part of the $l=2$, $m=2$ strain extracted from different radii
    or different resolutions ($\Delta x_{\text{finest}} = \si{\admmass}\slash65$ and
    $\Delta x_{\text{finest}} = \si{\admmass}\slash81$ respectively) from the simulation
    with $\lambda^{+}_{-} = 0.3$. The three waveforms agree well, indicating that the
    error due to the finite resolution or the finite extraction radius is
    small.}
\label{fig:comp-res}
\end{figure}

In Fig.~\ref{fig:comp-res}, we report how resolution and extraction radii affect
the resulting waveform. The differences between the various signals are small,
indicating the error due to the finite resolution and finite extraction radius
does not affect the results presented in this paper.

\section{Alternative formulations}
\label{sec:failed-attempts}

In our efforts to improve and stabilize our simulations, we tested two additional
formulations for the evolution of the electromagnetic fields. We found that
improvements arising from either of these formulations is subdominant compared
to the role of the dissipation (see, Sec.~\ref{sec:achieving-long-term}).

First, we followed~\cite{Farris2012} and implemented a generalized Lorenz
condition. If $A_{a}$ is the electromagnetic four-potential, the Lorenz gauge is
$\nabla_{a}A^{a} = 0$, while its generalized version is $\nabla_{a}A^{a} = \xi
n_{a} A^{a}$, with $\xi$ damping parameter. In~\cite{Farris2012}, it was found
that with a suitable choice of $\xi$, this condition reduces spurious gauge
modes that arise from interpolation at the refinement level boundaries. While
this is important for general-relativistic magneto-hydrodynamic simulations in
which matter crosses refinement levels, in our evolutions we did not find
significant improvements.

A second formulation we tested was motivated by the parabolized
Arnowitt-Deser-Misner formalism of general
relativity~\cite{Paschalidis:2007ey,Paschalidis:2007cp}. This consists of adding
an extra parabolic term to the evolution of the electric field, which is 0 when
the Gauss constraint is satisfied,
\begin{equation}
  \label{eq:rhs-E}
  \partial_{t} E^{i} = \cdots + \epsilon_{E} \gamma^{ij} \partial_{j} C_{E}\,.
\end{equation}
with $C_{E}$ is the Gauss constraint and $\epsilon_{E}$ strength of the parabolic term.
With this modification, the evolution of the constraint looks like
\begin{equation}
  \label{eq:rhs-C}
  \partial_{t} C_{E} = \cdots + \epsilon_{E} \gamma^{ij} \partial_{i} \partial_{j} C_{E}\,,
\end{equation}
which is parabolic diffusion operator. The indented result is to further
dissipate violations of the constraint from perturbations with high wave number.
In our tests, this formulation did not result in noticeable improvements over
our new method for setting the Kreiss-Oliger dissipation parameter presented in
Sec.~\ref{sec:achieving-long-term}.

\section{Newman-Penrose scalars in flat spacetime as function of electric and
  magnetic fields}
\label{sec:electr-newm-penr}

In this Appendix, we provide expressions for the electromagnetic Newman-Penrose
$\Phi_{0}$, $\Phi_{1}$, and $\Phi_{2}$ in terms of the electric and magnetic
fields in (asymptotically) flat spacetime. We consider both coordinate and
orthonormal bases. These expressions can be used to quickly compute the
Newman-Penrose scalars for a given electromagnetic field. (For a similar
discussion, see Appendix A in~\cite{Brennan2013} noting that a different
convention is used for the normalization of the tetrad.)

In flat spacetime and given the spherical coordinates $(r,\theta,\varphi)$,
consider the coordinate basis $(\vec{\partial}_{{r}},\vec{\partial}_{{\theta}},
\vec{\partial}_{{\phi}})$, and an orthonormal basis
$(\vec{e}_{\hat{r}},\vec{e}_{\hat{\theta}}, \vec{e}_{\hat{\phi}})$ so that for a
vector $\vec{v}$ we have that
\begin{subequations}
  \label{eq:E-B-orthonormal}
  \begin{align}
    \vec{v} &= v^{\hat{r}} \vec{e}_{\hat{r}} + v^{\hat{\theta}} \vec{e}_{\hat{\theta}}
    + v^{\hat{\varphi}} \vec{e}_{\hat{\varphi}}\,, \\
    \vec{v} &= v^{r} \vec{\partial}_{r} + v^{\theta} \vec{\partial}_{\theta}
    + v^{\hat{\varphi}} \vec{\partial}_{\varphi}\,,
  \end{align}
\end{subequations}
with
\begin{subequations}
    \label{eq:normaliz-vec}
  \begin{align}
    v^{\hat{r}} & = v^r\,,          & v_{\hat{r}}  &= v_r \,, \\
    v^{\hat{\theta}} & = r v^\theta\,,        & v_{\hat{\theta}}  &= \frac{1}{r} v_\theta \,, \\
    v^{\hat{\varphi}} & = r \sin \theta v^\varphi\,, & v_{\hat{\varphi}}  &= \frac{1}{r\sin \theta} v_\varphi \,.
  \end{align}
\end{subequations}
The electromagnetic field strength can be written in terms of the orthonormal
tetrad components of the electric and magnetic fields as follows
\begin{equation}
  \label{eq:Fmunu-NP-Mink}
  F_{ab} =
  \begin{pmatrix}
    0                    & -E_{\hat{r}}         & -r E_{\hat{\theta}}          & -r \sin \theta E_{\hat{\varphi}}  \\
    E_{\hat{r}}          & 0                    & r B_{\hat{\varphi}}           & -r \sin \theta B_{\hat{\theta}}  \\
    r E_{\hat{\theta}}        & -r B_{\hat{\varphi}}       & 0                       & r^2 \sin \theta B_{\hat{r}} \\
    r \sin \theta E_{\hat{\varphi}} & r \sin \theta B_{\hat{\theta}} & -r^2 \sin \theta B_{\hat{r}} & 0                      \\
  \end{pmatrix}
  \,.
\end{equation}

The Newman-Penrose scalars are, as computed by
Eqs.~\eqref{eq:NP-scalars-EM} and using Eq.~\eqref{eq:Fmunu-NP-Mink}
\begin{subequations}
  \label{eq:NP-scalars-EM-explicit}
  \begin{align}
    \Phi_0 & = \frac{1}{2} \left( -E_{\hat{\theta}} + B_{\hat{\varphi}} -i(E_{\hat{\varphi}} + B_{\hat{\theta}} )\right)  \,, \\
    \Phi_1 & = \frac{1}{2} \left( E_{\hat{r}} + iB_{\hat{r}} \right)  \,, \\
    \Phi_2 & = \frac{1}{2} \left( E_{\hat{\theta}} + B_{\hat{\varphi}} -i(E_{\hat{\varphi}} - B_{\hat{\theta}}) \right) \,.
  \end{align}
\end{subequations}
Alternatively, expressing Eqs.~\eqref{eq:NP-scalars-EM-explicit} in terms of
the vector components in the coordinate basis:
\begin{subequations}
  \label{eq:NP-scalars-EM-explicit-coordinate}
  \begin{align}
    \Phi_0 & = \frac{1}{2} \left[ -\frac{E_{\theta}}{r} + \frac{B_{\varphi}}{r\sin \theta} -i\left( \frac{E_{\varphi}}{r\sin \theta} +\frac{B_{\theta}}{r} \right)\right]  \,, \\
    \Phi_1 & = \frac{1}{2} \left( E_{r} + iB_{r} \right)  \,, \\
    \Phi_2 & = \frac{1}{2} \left[ \frac{E_{\theta}}{r} + \frac{B_{\varphi}}{r\sin \theta} -i\left(\frac{E_{\varphi}}{r\sin \theta} -  \frac{B_{\theta}}{r}\right) \right] \,.
  \end{align}
\end{subequations}
For completeness, we also report the null tetrad of
Eq.~\eqref{eq:null-tetrad} is (in the coordinate basis)
\begin{subequations}
  \label{eq:null-tetrad-coordinate}
  \begin{align}
    k^a          & = \frac{1}{\sqrt{2}} \left(1,-1,0,0\right) \,,  \\
    l^a          & = \frac{1}{\sqrt{2}} \left(1,1,0,0\right)\,,  \\
    m^a          & = \frac{1}{\sqrt{2}} \left(0,0,\frac{1}{r},\frac{i}{r\sin\theta}\right)\,, \\
    {\conj{m}}^a & = \frac{1}{\sqrt{2}} \left(0,0,\frac{1}{r},-\frac{i}{r\sin\theta}\right)\,.
  \end{align}
\end{subequations}

\section{Michel's solution}
\label{sec:michel-solution}

Michel's rotating magnetic monopole solution~\cite{Michel1973} is a simple model
for pulsar and black hole magnetospheres (for a rigorous description, see
~\cite{Gralla2014}, noting that Heaviside-Lorentz units are used, and many
formulas differ by a factor of $4\pi$ with what is reported here). Our interest
in Michel's monopole is motivated by the fact that it is a simple solution with
stationary flow of energy and angular momentum in flat spacetime that can be
used to test Eq.~\eqref{eq:flux-L} and its numerical implementation.

In an orthonormal basis, the non-zero components of electric and magnetic fields
of Michel's monopole are~\cite{Gralla2014}
\begin{subequations}
  \label{eq:E-B-Michel}
  \begin{align}
    E_{\hat{\theta}} & = B_{\hat{\varphi}}  = -\frac{q}{r}\omega \sin \theta \,, \\
    B_{\hat{r}} & = \frac{q}{r^2} \label{eq:Br-michel} \,.
  \end{align}
\end{subequations}
A four-vector potential that produces such configuration is given by
\begin{equation}
  \label{eq:A-michel}
  A_{\hat{\mu}} = \left(q\omega \cos \theta, 0, - \frac{q}{r} \omega \sin \theta, q \tan \frac{\theta}{2} \right)\,.
\end{equation}
Notice that $A_{\hat{\varphi}} \to + \infty$ when $\theta \to \pi$, but the resulting magnetic field
is well-behaved, as shown in Eq.~\eqref{eq:Br-michel}.

We can compute the Newman-Penrose scalars $\Phi_1$ and $\Phi_2$ with
Eqs.~\eqref{eq:NP-scalars-EM-explicit}:
\begin{subequations}
  \label{eq:NP-scalars-Michel}
  \begin{align}
    \Phi_1 & = \frac{i}{2} \frac{q}{r^2} \,, \\
    \Phi_2 & = -\frac{q}{r}\omega \sin \theta \,.
  \end{align}
\end{subequations}
As expected, $\Phi_{1}$ and $\Phi_{2}$ follow the peeling behavior of
Eqs.~\eqref{eq:NP-scalars-EM-falloff}. Next, we can use
Eq.~\eqref{eq:energy-em} to compute the total power that crosses a sphere
or radius $r$
\begin{eqnarray}
  \label{eq:power-Michel}
  \frac{\d E_{\text{EM}}}{\d t} =  \frac{2}{3} q^2 \omega^2\,.
\end{eqnarray}
We can also compute the flux of angular momentum along the $z$ direction
radiated with Eq.~\eqref{eq:flux-L}
\begin{eqnarray}
  \label{eq:flux-L-Michel}
  \frac{\d L^z_{\text{EM}}}{\d t} =  \frac{2}{3} q^2 \omega\,.
\end{eqnarray}
Hence, we find the relation
\begin{equation}
  \label{eq:omega-relation-Michel}
  \frac{\d E_{\text{EM}}}{\d t} =  \omega  \frac{\d L^z_{\text{EM}}}{\d t} \,.
\end{equation}
This is a well-known relation for pulsar and black hole magnetospheres.

\bibliographystyle{apsrev4-2} %
\bibliography{comparison-charged-bhs,einsteintoolkit}

\end{document}